# Photonic force optical coherence elastography for three-dimensional mechanical microscopy

Nichaluk Leartprapun[1], Rishyashring R. Iyer[1], Gavrielle R. Untracht[1,a], Jeffrey A. Mulligan[2], and Steven G. Adie[1,*]

[1] Meinig School of Biomedical Engineering, Cornell University, Ithaca, New York 14853

[2] School of Electrical and Computer Engineering, Cornell University, Ithaca, New York 14853

[a] Current affiliation: Optical and Biomedical Engineering Laboratory, School of Electrical, Electronic and Computer Engineering, The University of Western Australia, Perth, Western Australia 6009

* Corresponding author: sga42@cornell.edu

## Abstract

Optical tweezers are an invaluable tool for non-contact trapping and micro-manipulation, but their ability to facilitate high-throughput volumetric microrheology of biological samples for mechanobiology research is limited by the precise alignment associated with the excitation and detection of individual bead oscillations. In contrast, radiation pressure from a low numerical aperture optical beam can apply transversely localized force over an extended depth range. We propose photonic force optical coherence elastography (PF-OCE), leveraging phase-sensitive interferometric detection to track sub-nanometre oscillations of beads, embedded in viscoelastic hydrogels, induced by modulated radiation pressure. Since the displacements caused by ultra-low radiation-pressure force are typically obscured by absorption-mediated thermal effects, mechanical responses of the beads were isolated after independent measurement and decoupling of the photothermal response of the hydrogels. Volumetric imaging of bead mechanical responses in hydrogels with different agarose concentrations by PF-OCE was consistent with bulk mechanical characterization of the hydrogels by shear rheometry.

Optical manipulation has had a revolutionary impact in the biological and nanoscale sciences[1-3]. Developed by Ashkin and co-workers[4], optical tweezers (OTs) have enabled the manipulation of biological systems at the molecular-to-cellular scale. This has led to many seminal studies, including measurement of the elastic properties of bacterial flagella[5], direct observation of the movement and forces generated by molecular motors[6,7], the study of mechanotransduction pathways in living cells[8], and measurement of the mechanical properties and biophysical interactions of DNA[9,10]. OTs make use of the gradient force of a high-numerical aperture (NA) laser beam to achieve trapping and manipulation of micrometre-sized particles near the beam focus. Typical forces that can be achieved range from a piconewton to hundreds of piconewtons, which covers the range of forces associated with the biophysical processes of life at the molecular-to-cellular scale.

More than a decade before the invention of OTs, Ashkin demonstrated that low-NA laser beams could exert sufficient radiation-pressure forces to accelerate micrometre-scale dielectric particles along the beam path, or could be used to form a dual-beam trap based on counter-propagating beams[11]. Guck et al. adopted Ashkin's original dual-beam trapping configuration to develop the 'optical stretcher' for the study of cell mechanics[12], and used it to study the mechanical properties of cancer cells[13] and human blood cells[14]. Radiation pressure has been utilized on other microfluidic platforms for nanoparticle sorting and chromatography[15,16]. On a much larger scale, 'light sailing' by radiation pressure has been proposed as a mode of space travel in the 'solar sail'[17] and the Breakthrough Starshot[18] initiatives. Compared to OTs, however, optical manipulation based on radiation pressure has led to fewer applications in the life sciences.

The rapidly growing field of mechanobiology[19,20] has opened up new opportunities for optical manipulation on the micrometre-to-millimetre scale. Over the last decade, mechanobiology research has uncovered the integral role that extracellular matrix (ECM) mechanical properties and biophysical cell-ECM interactions play in both physiological processes and disease, including the onset and progression of cancer[21-24], stem cell differentiation[25-27], morphogenesis[28], and wound healing[27]. Biophysical interactions play an integral role across spatial scales, from molecular processes at the nanoscale[6-10], to collective (emergent) behaviour at the micro-to-mesoscale[29-31]. At the micro-to-mesoscale, an important trend toward the use of 3D cell culture systems (cell behavior is different in 2D versus 3D)[24,32,33] is driving the need for new imaging approaches that can support volumetric microscopy of mechanical properties. Engineered biological systems are the cornerstone of discovery-based mechanobiology research since they enable the systematic study of dynamic biomechanical and biochemical interactions under controlled/repeatable conditions. AFM has been used as the 'go to' method in mechanobiology, but can only interrogate the sample surface. Laser tweezers-based active microrheology[34,35] (AMR) is a promising approach that utilizes OTs to induce and detect nanometre-scale displacements from micro-beads randomly distributed within 3D cell culture, but its volumetric throughput is limited by the precise 3D alignment of high-NA trapping and position detection beams to each probing bead, prior to actuating it with a transversely oscillating optical trap. Although simultaneous trapping and detection of multiple beads can be achieved with time-shared OTs[36] or holographic OTs[37,38], multiplexed manipulation of beads randomly distributed over a depth range of hundred micrometres or greater with OTs has not been demonstrated. Other emerging approaches include Brillouin microscopy[39,40] and optical coherence elastography[41-43] (OCE). However, none of the above methods have demonstrated the ability to support mechanical microscopy with both 3D cellular-level resolution and sufficient throughput to facilitate time-lapse imaging studies over millimetre-scale volumes in mechanobiology.

In order to address the unmet need for volumetric, time-lapse imaging of the mechanical properties of biological systems, we revisited Ashkin's original idea of using radiation pressure from a low-NA beam[11]—now as a potential mechanism to apply localized mechanical excitation to micro-beads embedded in aqueous biological media. The use of low-NA radiation pressure to apply transversely-localized axial force over an extended depth range is advantageous for volumetric data acquisition, but comes with the challenges that a low-NA beam exerts significantly lower force than high-NA OTs, and that historically, studies on isolating the effects of photonic radiation pressure have been hindered by accompanying photothermal responses[11,44,45]. We present photonic force OCE (PF-OCE), leveraging the interferometric displacement sensitivity of OCT to detect picometre-to-nanometre bead oscillations induced by modulated radiation pressure from a low-NA beam. We address the challenge of isolating radiation pressure effects within absorbing aqueous media via a linear model to decouple mechanical and photothermal responses, combined with a differential scattering approach. PF-OCE has the potential to provide a new platform for large-scale volumetric mechanical microscopy, with a spatial sampling that is statistically controlled by the distribution of the beads inside the medium. Such a capability may readily find applications in cell mechanics and mechanobiology research, for instance, by enabling the mapping of microscale spatial variations in extracellular matrix (ECM) mechanics for 3D traction force microscopy (TFM)[29,31,33,46].

**Principle of PF-OCE**

Based on excitation by a harmonically modulated low-NA beam (hereafter referred to as the PF forcing beam), PF-OCE measures the resulting oscillations of beads embedded in viscoelastic media induced by the PF forcing beam (the "mechanical response") by compensating for the accompanying absorption-mediated photothermal effects of the aqueous medium (the "photothermal response").

We begin with the theoretical basis for using harmonically modulated radiation pressure to induce oscillations of beads embedded in a viscoelastic medium. Generalizing Ashkin's simplified expression[11] to include the contribution of both photon scattering and absorption to the net change in linear momentum, the radiation-pressure force, $\mathbf{F_{rad}}$, exerted on a neutral particle by a weakly focused beam with optical power $P$, is given by:

$$\mathbf{F_{rad}} = \frac{(2q_s + q_a)n_{med}P}{c}\hat{\mathbf{z}}, \tag{1}$$

where $c$ denotes the speed of light in vacuum, $n_{med}$ denotes refractive index of the medium and $\hat{\mathbf{z}}$ denotes a unit vector pointing in the propagation direction of the forcing beam. The proportionality constants $q_a$ and $q_s$ define the fractions of incident photon momentum that are imparted to the bead in the direction $\hat{\mathbf{z}}$ as a result of absorption and backscattering, respectively. For a non-absorbing particle, such as the latex (polystyrene) beads used in Ashkin's experiments, the contribution of $q_a$ to the radiation-pressure force is neglected. The proportionality constant $q_s$ accounts for the effects of shape, size, and scattering cross-section of the bead in relation to the characteristics of the forcing beam (e.g.

wavelength and beam waist radius or NA), the refractive indices of the bead and medium, and the position of the bead in 3D space relative to the forcing beam.

Together with the restoring force from the viscoelastic medium, harmonically modulated **F**$_{\text{rad}}$ induces oscillatory motion of the bead. Oestreicher provided a theoretical model for the impedance of an oscillating sphere in a viscoelastic medium[47]. We inverted equation (18) in Oestreicher's paper[47] to obtain an expression for the oscillation amplitude of the bead as a function of bead radius, $a$, complex shear modulus, $G^*(\omega) = G'(\omega) + iG''(\omega)$, and mass density, $\rho$, of a linear viscoelastic medium, given by

$$u_0(\omega) = \frac{F_{\text{rad}}}{6\pi a |G_{\text{eff}}(\omega)|}, \tag{2a}$$

where

$$G_{\text{eff}}(\omega) = \frac{\rho a^2 \omega^2}{9} - G^*(\omega)\left[1 - i\sqrt{\frac{\rho a^2 \omega^2}{G^*(\omega)}}\right]. \tag{2b}$$

In equation (2a), $u_0$ describes the oscillation amplitude of the bead resulting from harmonically modulated radiation-pressure force with peak magnitude $F_{\text{rad}} = \|\mathbf{F}_{\text{rad}}\|$ and modulation (angular) frequency $\omega$. We observe that $u_0$ is directly proportional to the magnitude of radiation-pressure force but inversely proportional to high-order powers of the bead radius.

For non-absorbing beads embedded in an aqueous medium, the scattering-mediated radiation pressure exerted on the bead from the PF forcing beam is accompanied by the absorption-mediated photothermal response of the medium. Absorption-mediated responses form the basis of multiple functional imaging modalities. Photoacoustic tomography (PAT) uses short laser pulses to generate ultrasonic pressure waves from absorption-induced thermoelastic expansion[48]. High-power laser pulses have also been used to generate propagating surface acoustic waves caused by thermal expansion for elastography applications[49]. Absorption-induced optical path length (OPL) change, governed by thermo-optic effect and thermal expansion, allows photothermal OCT (PT-OCT) to detect the presence of chromophores in biological samples[50,51].

In order to understand and account for the effects of absorption on OPL, consider the case of a non-absorbing dielectric bead embedded at depth $L$ in a homogeneous absorbing medium with uniform (spatially invariant) refractive index $n_{med}$. The OPL to the bead measured by OCT is encoded in the phase of the complex OCT signal, given by $\Phi = (4\pi/\lambda) \cdot n_{med} L$. The OPL to the bead corresponds to the product $\text{OPL} = n_{med} L$. Both $n_{med}$ and $L$ can vary with a change in temperature via two different phenomena—$n_{med}$ via the thermo-optic effect and $L$ via thermal expansion. The OPL change with respect to the change in temperature can be expressed by a product rule of differentiation,

$$\frac{d\text{OPL}}{dT} = \frac{dn_{med}}{dT} L(T) + n_{med}(T) \frac{dL}{dT}. \tag{3}$$

The derivatives $dn_{med}/dT$ and $dL/dT$ are the thermo-optic coefficient and the thermal expansion coefficient of the medium, respectively. Equation (3) describes how the measured OPL change due to the change in temperature of the medium originates from both the change in refractive index and thermal expansion, and does not directly correspond to physical displacement of the bead.

Assuming that scattering and absorption are independent events within the context of PF-OCE, we model the measured bead OPL oscillation (the "total response"), $\Delta\text{OPL}_{tot}$, due to the modulated PF forcing beam as a linear combination of the complex mechanical response of the bead, $\Delta\text{OPL}_{mech}$, and the complex photothermal response of the medium, $\Delta\text{OPL}_{PT}$, given by

$$\Delta\text{OPL}_{tot}(\mathbf{r}, t, \omega) = \Delta\text{OPL}_{mech}(\mathbf{r}, t, \omega) + \Delta\text{OPL}_{PT}(\mathbf{r}, t, \omega) \tag{4a}$$

where $\Delta\text{OPL}_{mech}$ and $\Delta\text{OPL}_{PT}$ are given by

$$\Delta\text{OPL}_{mech}(\mathbf{r}, t, \omega) = A_{mech}(\mathbf{r}, \omega) e^{i(\omega t + \varphi_{drive} + \varphi_{mech}(\mathbf{r}, \omega))} \tag{4b}$$

and

$$\Delta \text{OPL}_{\text{PT}}(\mathbf{r}, t, \omega) = A_{\text{PT}}(\mathbf{r}, \omega) e^{i(\omega t + \varphi_{\text{drive}} + \varphi_{\text{PT}}(\mathbf{r}, \omega))}. \quad (4c)$$

Vector $\mathbf{r} = (x, y, z)$ denotes the spatial coordinates of each pixel on the OCT image and $\varphi_{\text{drive}}$ denotes the phase of PF forcing beam drive waveform at time $t = 0$. $A_{\text{mech}}$ and $\varphi_{\text{mech}}$ denote the amplitude and phase of the complex mechanical response, respectively. Likewise, $A_{\text{PT}}$ and $\varphi_{\text{PT}}$ denote the amplitude and phase of the complex photothermal response. The goal of PF-OCE is to isolate the complex mechanical response of the bead, which is dependent on the mechanical properties of the surrounding viscoelastic medium, from the measured total response by subtracting accompanying photothermal response of the medium (Fig. 1a).

## Theoretical simulations of mechanical and photothermal responses

In order to understand the effects of each design parameter (e.g. NA of the PF forcing beam and bead size) on the measured response, and to obtain an estimate of the expected magnitude of the mechanical and photothermal responses, we simulated the contributions of both radiation-pressure force and photothermal response to the total OPL oscillation of a polystyrene bead. Unless stated otherwise, all numerical results presented in this section were obtained from theoretical simulation assuming a Gaussian PF forcing beam with wavelength $\lambda = 976$ nm and waist radius $w_0 = 3.19$ μm, and a spherical bead with refractive index $n_{\text{bead}} = 1.58$. Refractive index of the medium was assumed to be $n_{\text{med}} = 1.34$ for biological hydrogels.

An accurate estimate of the backscattered (and absorbed) photon energy is critical to predict the magnitude of radiation-pressure force on a bead. Several approaches have been used to estimate $q_s$ and $q_a$ in equation (1). Among them, Generalized Lorenz-Mie Theory (GLMT)[52] is applicable for estimating $F_{\text{rad}}$ from a focused Gaussian beam on a bead of arbitrary shape and size. Our MATLAB implementation of GLMT showed that the normalized radiation-pressure force (i.e., $F_{\text{rad}}$ per unit power), $\bar{F}_{\text{rad}}$, from the PF forcing beam on a bead was on the order of 0.2-0.3 pN/mW at the focal plane when the focal spot size of the beam was comparable to the bead diameter (Fig. 1b); the force was lower as the two deviated. Additionally, $\overline{F}_{\text{rad}}$ became larger as the beam waist and the bead both increased in size while maintaining comparable diameters, which can be attributed to a higher backscattering cross-section of a larger bead. The resulting normalized bead oscillation amplitude, $\overline{u}_0$, similarly decreased as the diameters of the beam waist and the bead deviated from each other (Fig. 1c) since $u_0$ is directly proportional to $F_{\text{rad}}$ (Equation 2a). However, whereas $\overline{F}_{\text{rad}}$ increased monotonically with both beam waist and bead diameter (when the two dimensions are comparable), $\overline{u}_0$ decreased beyond a certain point (Fig. 1c) because the oscillation amplitude is also inversely proportional to $a$ (Equation 2a, b).

In order to simulate the photothermal response, we solved the heat transfer equation to estimate the change in temperature of the medium due to absorption and then modified a theoretical model given for PT-OCT by Lapierre et al.[51] to estimate the resulting cumulative OPL change for a general case with spatially varying $n_{\text{med}}$ and $T$. Detailed descriptions of the theoretical model and parameters used for the simulation can be found in Supplementary Method 1 and Supplementary Table 1 respectively. In an aqueous medium, the normalized change in temperature per unit power, $\overline{\Delta T}$, due to absorption by water molecules was simulated to be on the order of $10^{-2}$ K/mW after 50 ms of continuous exposure to the PF forcing beam (Fig. 1d). The resulting normalized cumulative OPL change, $\overline{\Delta \text{OPL}}$, due to the photothermal effects were on the order of 0.5 nm/mW (Fig. 1e), approximately an order of magnitude larger than $\overline{u}_0$ measured for the same beam parameters ($\lambda = 976$ nm, $w_0 = 3.19$ μm, and $n_{\text{med}} = 1.34$) (Fig. 1c, e).

These simulations assume that the OPL change induced by the photothermal effects of a PF forcing beam on a given medium is unaffected by the presence or size of the beads. On the other hand, both the radiation-pressure force and the resulting bead oscillation amplitude for a given medium and PF forcing beam can vary by an order of magnitude depending on the size of the bead alone. This provides a guide for designing experimental conditions that affect the mechanical response without disturbing the photothermal response, forming the basis for the isolation of bead mechanical response from the measured total response (described in Section: Isolation of bead mechanical response).

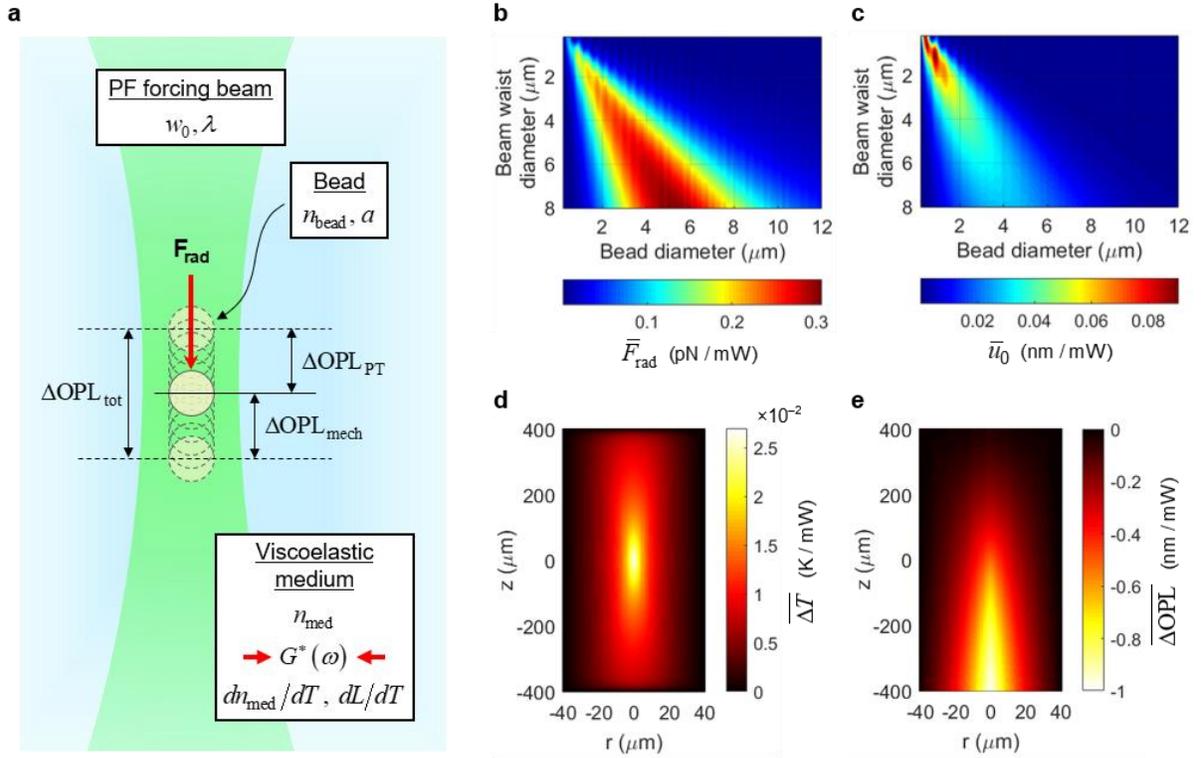

**Figure 1. Underlying principle of PF-OCE for AMR in viscoelastic biological media, and theoretical prediction of bead responses resulting from radiation-pressure force and photothermal effects. a**, Cartoon illustration of the working principle of PF-OCE depicting mechanical excitation by modulated radiation-pressure force, $\mathbf{F}_{rad}$, from a low-NA beam and various factors, associated with the PF forcing beam, the bead, and the viscoelastic medium, that affect the measurement of change in OPL. **b**, Magnitude of normalized forward radiation-pressure force, $\overline{F}_{rad}$, from a Gaussian beam ($\lambda = 976$ nm) on a non-absorbing spherical bead embedded in a medium (refractive indices $n_{bead} = 1.58$ and $n_{med} = 1.34$) as a function of bead diameter, $2a$, and beam waist diameter, $2w_0$, obtained from simulation based on GLMT. **c**, Normalized (physical) bead oscillation amplitude, $\overline{u}_0$, resulting from harmonically-modulated radiation-pressure force with peak amplitude $\overline{F}_{rad}$ and modulation frequency $\omega = 2\pi(20\text{ Hz})$ for a viscoelastic medium with $G^*(\omega) = 250 + i4$ (Pa). **d**, Map of normalized change in temperature, $\overline{\Delta T}$, of an aqueous medium after 50 ms of exposure to a Gaussian beam ($\lambda = 976$ nm, $w_0 = 3.19$ μm). **e**, Map of cumulative optical path length change, $\overline{\Delta \text{OPL}}$, induced by $\overline{\Delta T}$ based on a PT-OCT model (Supplementary Method). Note that **e** represents OPL and not physical distance. In **d** and **e**, $r$ and $z$ denote the radial and axial (depth) coordinates, defined w.r.t. the focus of the beam, respectively.

**Experimental setup and data acquisition**

To enable simultaneous mechanical excitation by radiation pressure from the PF forcing beam and detection of resulting OPL oscillations by phase-sensitive OCT, we combined the PF forcing beam with the sample arm beam of a spectral-domain (SD)-OCT system via a free-space beam control module (BCM) and a dichroic filter (Fig. 2a). We selected the PF forcing beam wavelength of 976 nm for its low to negligible cytotoxicity reported in previous OTs studies[53]. In this configuration, both the OCT and the PF forcing beams were collinearly scanned in a raster pattern by the same galvanometer, and focused by the same objective lens. The BCM was adjusted to ensure that both beams were co-aligned and focused to the same position in 3D space (Supplementary Method 4), and the overall PF optical system was designed to minimize optical aberrations. The waist radius of the PF forcing beam was measured to be 3.19 µm (NA ~0.1) at the focal plane (Supplementary Figure 1 and Supplementary Method 3). Although the theoretical simulation suggests that a PF forcing beam with smaller waist radius (when paired with comparable bead size) is optimal for maximizing the bead oscillation amplitude (Fig. 1b), we chose not to increase the NA of the PF forcing beam beyond 0.1 to ensure that radiation-pressure force would be applied over an extended depth range for high-throughput volumetric measurement.

We used agarose hydrogel as an example of a viscoelastic substrate used in cell imaging applications. The sample consisted of agarose hydrogel inside a confined glass chamber (Fig. 2a). Mechanical properties of the hydrogel were varied by changing agarose polymer concentration, from 0.2-0.5 %w/w (hereafter referred to as 0.2-0.5% hydrogels) in distilled water to achieve shear modulus in the range of 0.1-1 kPa[54]. Polystyrene beads with mean particle diameter of 3 µm were added to the hydrogels to serve as non-absorbing scattering particles for oscillation by radiation pressure. Under these experimental conditions, we expect $\bar{u}_0$ to be approximately 0.03 nm/mW.

To enable 3D volumetric measurements of OPL oscillations of the 3-µm beads, we adopted a 3D BM-mode acquisition scheme, which allows continuous beam-scanning acquisition along the fast axis (B-scan) while also supporting OPL tracking at each spatial location over time (M-scan) (Fig. 2b). For the experiments presented here, a total of 6,144 frames (B-scans) were acquired consecutively at 200 frames/s at the same slow-axis position. Meanwhile, the PF forcing beam power was externally modulated by a continuous 20-Hz sinusoidal drive waveform from a function generator. Thus, the OPL oscillation at each spatial voxel was tracked at 6,144 instances in time over 614 modulation cycles of the PF forcing beam.

The BM-mode acquisition scheme has three key implications for the implementation of PF-OCE in biological systems such as in live-cell imaging studies. Firstly, the fast-axis beam-scanning effectively resulted in a pulse-train mechanical excitation on each of the 3-µm beads instead of a continuous sinusoidal waveform provided by the function generator (Supplementary Figure 3). As a B-scan was acquired, the PF forcing beam would dwell on each of the 3-µm beads for ~67 µs (corresponding to 4 A-scans), after which the bead received no force until the PF forcing beam scanned over the bead again in the next frame. In other words, the actual excitation on each bead is a frequency comb with a 20-Hz fundamental frequency and other higher order harmonics. This implies that the frequency-dependent response of the medium must be accounted for when quantitatively reconstructing mechanical properties of the medium from the bead responses; this subject will be addressed in a future manuscript. Under this type of excitation, the time-averaged optical power imparted on each 3-µm bead by the PF forcing beam was only 0.3 mW for a peak power of 112 mW, as opposed to 56 mW that would have been expected from a continuous sinusoidal excitation with the same peak power. Although this outcome is expected to result in a lower bead oscillation amplitude compared to the continuous excitation case, the 2 orders of magnitude reduction in the time-averaged optical power imparted on the sample is beneficial for biological studies where cell viability is a concern.

Secondly, there is a trade-off between acquisition speed and OPL oscillation measurement sensitivity. In the shot-noise limit, the sensitivity of OPL oscillation amplitude measurement by OCT is approximately inversely proportional to the square-root of the number of BM-mode frames acquired per slow-axis position[55,56] (Supplementary Method 5). Prioritizing sensitivity over speed, we acquired up to 6,144 frames per slow-axis position and achieved OPL oscillation amplitude noise floor of 75 pm, approximately 20 pm above the theoretical shot-noise limit (Supplementary Method 5), for OCT signals with signal-to-noise ratio (SNR) of 25 dB (Supplementary Figure 2). This acquisition scheme requires at least 5 minutes to acquire OPL oscillation data over a transverse field-of-view (FOV) of 200 µm × 10 µm at a spatial sampling of 1 µm/pixel; this spatial sampling density ensures each 3-µm bead was sampled multiple times along the fast and slow axes. Alternatively, the acquisition time can be shortened without sacrificing the displacement sensitivity by increasing the frame rate; this can be achieved with a resonant scanner or other high-speed beam scanning options[57].

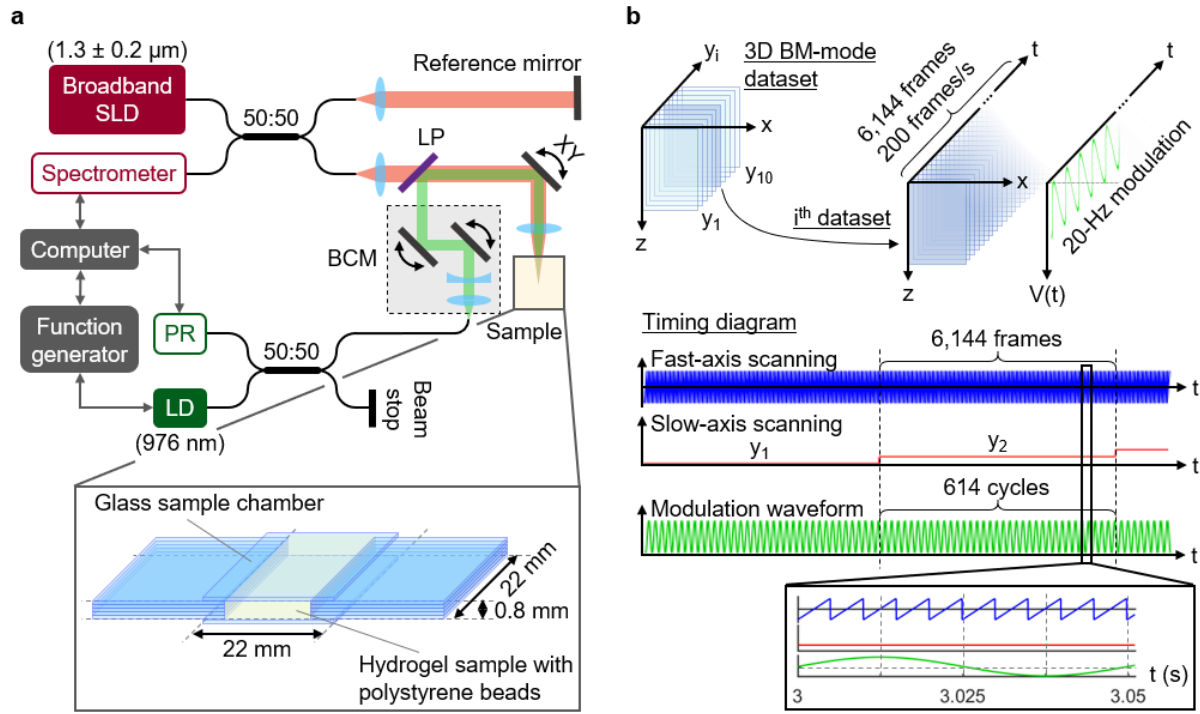

**Figure 2. Optical setup and data acquisition scheme for volumetric AMR with PF-OCE. a**, Optical setup consisted of an SD-OCT system and a PF forcing beam combined in free space with the OCT sample arm beam. A function generator provided external modulation of the power of the PF forcing beam. **b**, Illustrations of 3D BM-mode dataset and associated timing diagram for the beam scanning (200-Hz frame rate) and modulation waveform (20-Hz modulation frequency) from the function generator for a 3D BM-mode acquisition. SLD: superluminescent diode, LD: laser diode, PR: photo receiver, LP: long-pass dichroic filter, BCM: beam control module, XY: two-axis galvanometer.

**Isolation of bead mechanical response**

Based on the linear model in equation (4a-c), the mechanical response of the 3-μm beads can be isolated from the measured total response by subtracting the photothermal response of the medium. This approach requires availability of a reliable estimate of the photothermal response. In principle, the complex photothermal response in a uniformly absorbing medium may be theoretically obtained from the model of the absorption-mediated OPL change in PT-OCT[51]. However, we were not able to ascertain the accuracy of the theoretical simulation (Supplementary Method 1) under our experimental conditions due to the lack of available material properties for the agarose hydrogels used in our experiments as well as the added contribution of the confined glass chamber in the experimental setup to the heat transfer process and thermal expansion model (Supplementary Discussion 1). Alternatively, under the premise that weak scattering signals from the medium would produce adequate OCT signals for OPL measurements but would not be sufficient to produce detectable mechanical response induced by scattering-mediated radiation-pressure force, a differential scattering approach could be employed wherein the complex photothermal response may be measured experimentally from weak scatterers in the sample. These weak scatterers thus act as 'reporters' of the photothermal response, without producing measurable displacements resulting from photon momentum transfer.

In implementing this differential scattering approach, we leveraged the size-dependence of backscattering intensity[3] and added 0.1-μm polystyrene beads to the sample to provide weak background scattering signals for measuring the photothermal response. The observed OCT scattering intensity from the 0.1-μm beads was approximately 3 orders of magnitude lower than the 3-μm beads, which is indicative of the ratio of the photon momentum transfer to the beads within the excitation volume. Based on the OCT scattering intensity and accounting for the average number of beads per PF forcing beam excitation volume, we estimated approximately 5 orders of magnitude lower $\bar{F}_{\text{rad}}$ per bead for the 0.1-μm beads compared to the 3-μm beads (Supplementary Method 2). Given their smaller size, this difference would result in approximately 4 orders of magnitude smaller $\bar{u}_0$ for the 0.1-μm beads, supporting the assumption that the mechanical response of the 0.1-μm beads was negligible. The estimation based on observed difference in the OCT scattering intensity from the two bead sizes agreed with the theoretical predictions based on GLMT[52] and Oestreicher's model[47] (Supplementary Method 2).

The process to isolate the mechanical responses of the 3-µm beads is summarized in a flow chart (Fig. 3). In order to obtain a reliable estimate of the photothermal response, it was important to account for the consequences of the low SNR of OCT signals[55,56] from the weakly scattering 0.1-µm beads. The low OCT SNR (<12 dB) of the 0.1-µm beads resulted in relatively large OCT phase noise on the measured OPL oscillations (i.e., the raw photothermal response). To reduce the contribution of individual OCT phase measurement errors, we calculated the depth-dependent photothermal response (with amplitude $A_{\mathrm{PT}}(z)$ and phase $\varphi_{\mathrm{PT}}(z)$) by performing curve-fitting of the amplitude and phase of the raw photothermal response as a function of depth (Fig. 3), using the theoretical simulation of the absorption-mediated OPL change[51] for the amplitude and a cubic polynomial function for the phase (Method 4). We assumed that water was the only absorber in the sample and that water concentration was transversely uniform across the sample, resulting in a photothermal response that is dependent on depth alone. This combined experimental and theoretical approach yielded estimates of the depth-dependent photothermal response from the fitted curves with uncertainties of approximately ±0.7 nm for $A_{\mathrm{PT}}(z)$ and ±0.5 rad for the $\varphi_{\mathrm{PT}}(z)$ (Supplementary Figure 5). The mechanical responses of the 3-µm beads, $\Delta\mathrm{OPL}_{\mathrm{mech}}$, were subsequently isolated from the measured total responses, $\Delta\mathrm{OPL}_{\mathrm{tot}}$, by subtracting the depth-dependent photothermal responses (values taken from the best-fit curves at the depths of the 3-µm beads) from the total responses (Method 4).

The approach described here relies on four fundamental assumptions. Firstly, we assumed that absorption events and scattering events that occurred in the sample were independent and separable. This assumption must hold for the linear model in equation (4a-c) to be valid. Secondly, we assumed that the mechanical response of the 0.1-µm beads was negligible. This assumption was supported by both theoretical predictions and experimental observation of OCT scattering intensity (Supplementary Method 2). Thirdly, we assumed that the photothermal response was transversely uniform. Under this assumption, the uncertainty from the curve fits may impose a depth-dependent systematic error that affects the accuracy of the isolated amplitude and phase of the mechanical responses relative to their true values, but does not degrade the precision for distinguishing microscale variations in the mechanical responses within a sample. However, if the photothermal response were to have a transverse variation that was unaccounted for, these uncertainties could also impose a random error that adversely affects the ability of PF-OCE to distinguish relative differences in the mechanical responses within a sample. Lastly, we assumed the photothermal response measured on the 3-μm beads was equivalent to the photothermal response measured on the 0.1-μm beads at the same depth. In other words, we assumed that any perturbations of the photothermal response specifically due to the presence of the 3-μm beads were negligible.

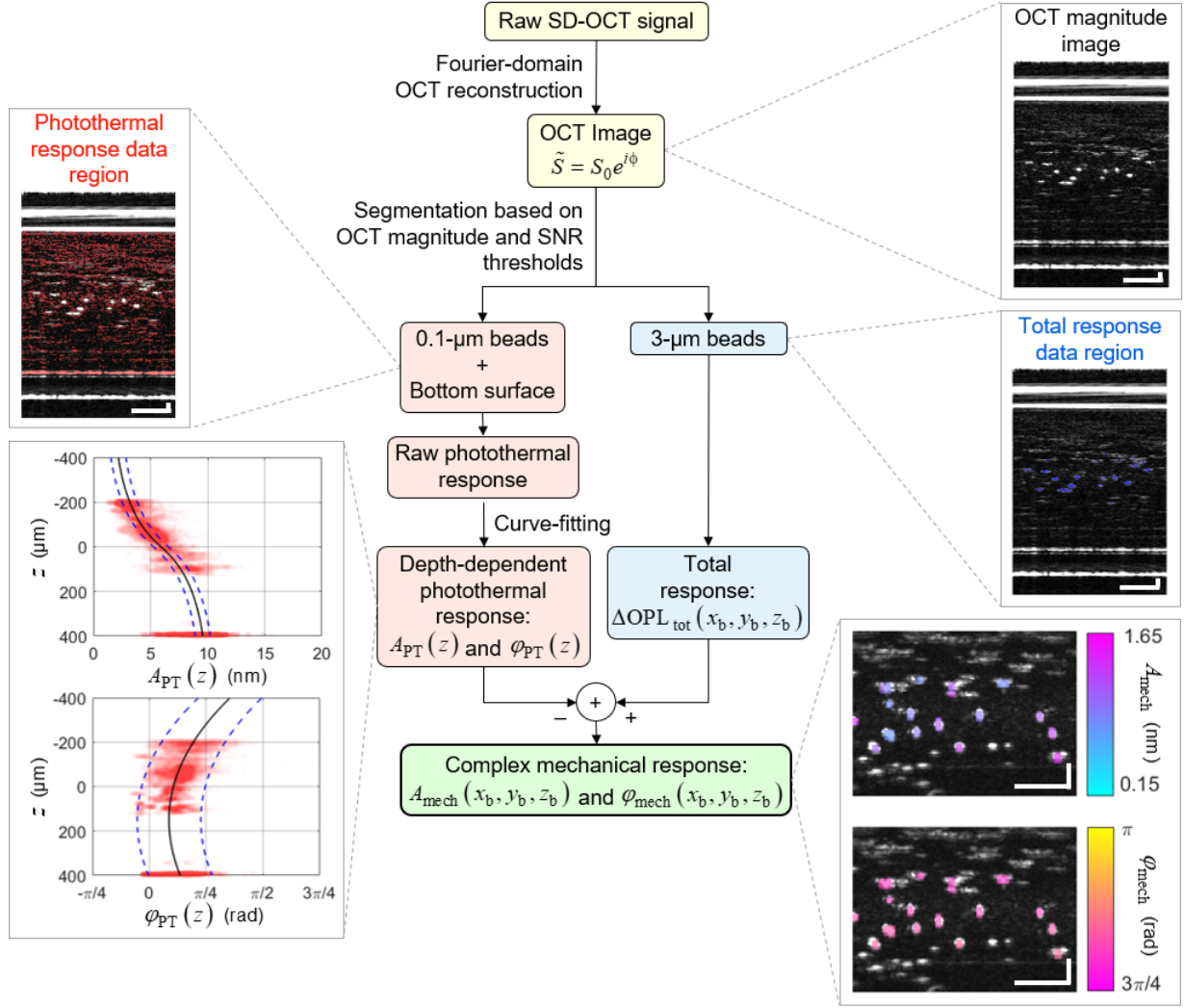

**Figure 3. Data processing flow chart outlining key steps to isolate the mechanical response of 3-μm beads from the measured total response by estimating and compensating for the photothermal response of the surrounding medium.** The OCT image was segmented into total response data region (3-μm beads) and photothermal response data region (0.1-μm beads and bottom surface of the sample chamber) with magnitude and SNR thresholds (Supplementary Table 2 provides exact values of the thresholds used in the presented experiments). Magnitude and phase of raw photothermal response data were fit to a theoretical photothermal response curve and a cubic polynomial function, respectively, to obtain depth-dependent photothermal response, given by amplitude $A_{\text{PT}}(z)$ and phase $\varphi_{\text{PT}}(z)$. The mechanical response of each 3-μm bead, given by amplitude $A_{\text{mech}}(x_b, y_b, z_b)$ and phase $\varphi_{\text{mech}}(x_b, y_b, z_b)$ where $(x_b, y_b, z_b)$ are the set of pixel coordinates corresponding to each 3-um bead, was isolated after subtracting the photothermal response at corresponding depths from the measured total response. Example images from a 0.4% agarose hydrogel dataset are provided at each key processing step.

**Volumetric AMR in agarose hydrogels**

The mechanical responses of the 3-μm beads were measured in four agarose hydrogel samples with different mechanical properties and compared to the bulk characterization of the hydrogels by shear rheometry (Supplementary Method 6). The complex total responses measured on the 3-μm beads, the fitted photothermal responses at the corresponding depths, and the isolated mechanical responses are displayed on the complex plane (Fig. 4a). We qualitatively observed that $\Delta\text{OPL}_{\text{tot}}$ and $\Delta\text{OPL}_{\text{PT}}$ had comparable magnitudes, and were distributed over the same quadrant of the complex plane for all agarose concentrations. In contrast, $\Delta\text{OPL}_{\text{mech}}$ was approximately 4 times smaller in magnitude and was in a different quadrant. These observations suggest that the measured total responses were dominated by contributions from photothermal effects, which can be explained by absorption of water molecules in the hydrogel samples at the PF forcing beam wavelength[58]. The fact that $\Delta\text{OPL}_{\text{PT}}$ and $\Delta\text{OPL}_{\text{mech}}$ are in different quadrants of the complex plane implies that the photothermal responses and the mechanical responses do not occur in-phase. Additionally, $\Delta\text{OPL}_{\text{mech}}$ followed a general trend of decreasing magnitude with increasing agarose concentration, consistent with our expectation that a higher agarose concentration would produce a stiffer hydrogel. This trend can be clearly observed when the median amplitude of the mechanical response (calculated from a set of spatial pixels corresponding to each 3-μm bead as described in Method 4) is overlaid on top of the OCT image of each sample (Fig. 4b). Similarly, a general trend as a function of agarose concentration could also be observed in the phase of the mechanical response, which approached $\pi$ phase delay as the concentration increased (Fig. 4c). In addition to mechanical contrast across samples, these maps also revealed the variability in $A_{\text{mech}}$ and $\varphi_{\text{mech}}$ of different beads within each sample, which may reflect microscale heterogeneity in the structural and mechanical properties of low-concentration agarose hydrogels[59-61]. For instance, beads with higher $A_{\text{mech}}$ within a sample could be those inside larger pores, diffusing in the fluid phase of the biphasic porous hydrogel, where as those with lower $A_{\text{mech}}$ could be trapped in the solid agarose polymer matrix[59-61]. Comprehensive discussions of sources of variability in PF-OCE measurements can be found in Supplementary Discussion 2.

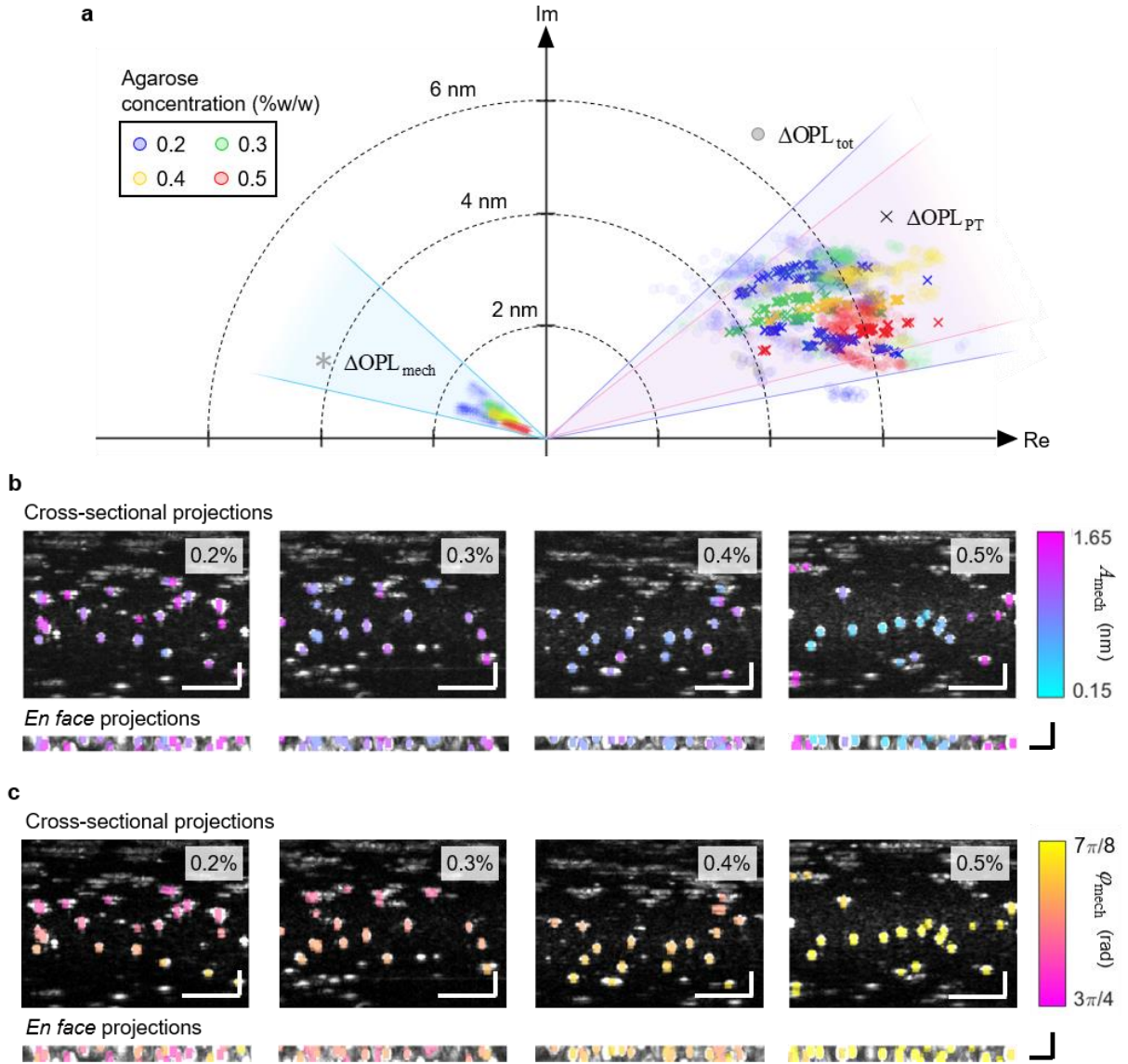

**Figure 4. PF-OCE measurements in four agarose hydrogel samples with different agarose concentrations.**
**a**, All measurements of $\Delta \text{OPL}_{\text{tot}}(x_b, y_b, z_b)$ (●), $\Delta \text{OPL}_{\text{PT}}(z_b)$ (✗) after curve-fitting, and $\Delta \text{OPL}_{\text{mech}}(x_b, y_b, z_b)$ (∗) displayed on the complex plane. **b**, $A_{\text{mech}}(x_b, y_b, z_b)$ from one 3D BM-mode dataset in each sample displayed as a map overlaid on cross-sectional and *en face* OCT image maximum intensity projections (grayscale). **c**, corresponding maps of $\varphi_{\text{mech}}(x_b, y_b, z_b)$ from the same datasets as in **b**. Scale bars: 50 μm (white, cross-sectional projections) and 20 μm (black, *en face* projections).

For quantitative comparisons to standard shear rheometry, $A_{\text{mech}}$, $A_{\text{PT}}$, $\varphi_{\text{mech}}$, and $\varphi_{\text{PT}}$ measured by PF-OCE are displayed as box plots next to the magnitude of complex shear modulus, $|G^*|$, and phase delay, $\varphi_{\text{rhe}} = \tan^{-1}(G''/G')$, measured by shear rheometry (Fig. 5). The total response is omitted here but can be found in Supplementary Figure 6. Complete results from shear rheometry can be found in Supplementary Figure 7. We found a statistically significant ($p_C \ll 0.05$, see Method 5 for statistics) monotonically decreasing trend in $A_{\text{mech}}$ versus agarose concentration (Fig. 5a). This behaviour agrees with progressively increasing $|G^*|$ of the hydrogels as the agarose concentration increased (Fig. 5c). In contrast, no significant trend ($p_C = 0.15$) or difference across agarose concentrations was observed for $A_{\text{PT}}$ (Fig. 5b). We also found a similar increasing trend ($p_C \ll 0.05$) in both $\varphi_{\text{mech}}$ and $\varphi_{\text{rhe}}$ towards $\pi$ as the agarose concentration increased (Fig. 5d, f). The phase delay approaching $\pi$ is consistent with the response of a predominantly elastic material excited above its damped natural frequency. In contrast, $\varphi_{\text{PT}}$ followed an opposite trend ($p_C \ll 0.05$) and was closer to 0 for all hydrogels (Fig. 5e). We note that $\varphi_{\text{mech}}$ was up to $\pi/4$ rad smaller than $\varphi_{\text{rhe}}$ for all concentrations. The discrepancies may reflect the differences between bulk responses of the hydrogels measured by shear rheometry and microscale mechanical responses measured by PF-OCE. Nevertheless, both rheometry and PF-OCE suggest that the hydrogels become more elastic as the agarose concentration increases, which is consistent with previous studies that reported decreases in porosity and influence of viscous drag (due to fluid flow through pores) at higher agarose concentrations[60,61]. Our results demonstrate that the isolated complex mechanical responses of the 3-µm beads from PF-OCE can be used to distinguish different viscoelastic properties of agarose hydrogels, whereas no significant trend that directly correlates to the mechanical properties was observed in the photothermal responses.

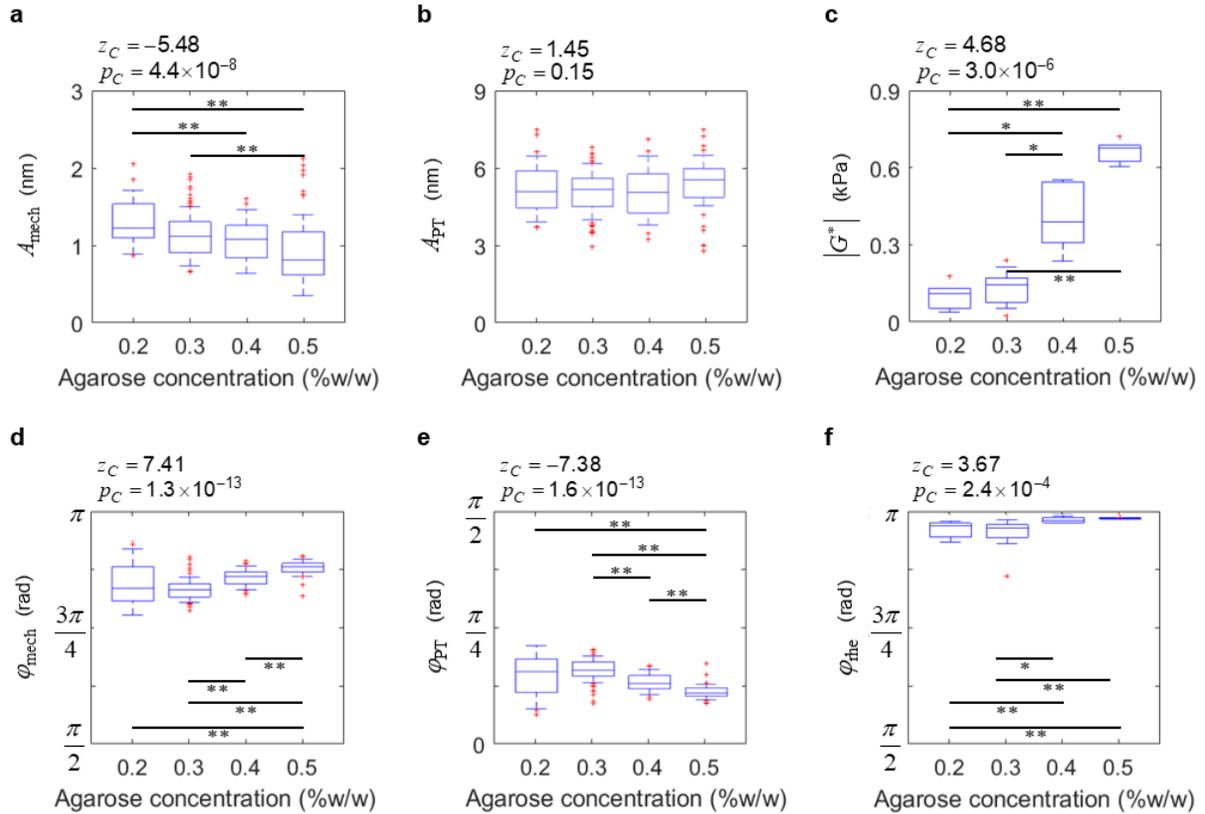

**Figure 5. Box plots comparing the complex mechanical and photothermal responses in agarose hydrogels to bulk mechanical characterization of the hydrogels by shear rheometry.** Magnitude and phase of **a**, **d**, mechanical response, and **b**, **e**, photothermal response compared to **c**, **f**, magnitude and phase of the complex shear modulus measured by parallel-plate shear rheometry. Plotted data from PF-OCE measurements represent median values obtained from each of the 3-μm beads within the FOV (total of N = 54, 74, 58, and 52 beads for 0.2%, 0.3%, 0.4%, and 0.5% agarose hydrogel samples, respectively). Rheometer measurements were taken on 3 separate samples, with 3 repetitions per sample, for each agarose concentration (total of N = 5, 9, 9, and 6 independent measurements of 0.2%, 0.3%, 0.4%, and 0.5% agarose hydrogel samples, respectively). Horizontal lines within boxes indicate median values, boxes denote interquartile ranges. Error bars span one standard deviation; data outside of this range are shown in red markers. Black bar and asterisks indicate a statistically significant difference between 2 agarose concentrations per Kruskal-Wallis test at $p<0.05$ (*) and $p<0.005$ (**) confidence levels. $z_C$ and $p_C$, respectively, denote normalized test statistic and associated p-value for Cuzick's test for trend across the four agarose concentrations, ordered from 0.2% to 0.5% w/w; thus, $z_C>0$ indicates an increasing trend while $z_C<0$ indicates a decreasing trend. A trend was considered statistically significant if $p_C<0.05$. Refer to Method 5 for details of the statistical analysis.

**Discussion**

The use of radiation pressure to induce bead oscillations has two key implications for the basis of PF-OCE. Firstly, the ultra-low radiation-pressure force from a low-NA beam results in picometre-to-nanometre bead oscillation amplitudes, which pushes the limit of interferometric axial displacement sensitivity of phase-sensitive OCT. Our results demonstrate an oscillation amplitude measurement sensitivity of ≤75 pm (for OCT SNR ≥25 dB), which, in practice, corresponds to the smallest detectable oscillation amplitude of 150 pm for a detection threshold of 3 dB. The current sensitivity was achieved after optimization of hardware synchronization and acquisition scheme to minimize galvanometer motion instability.

Secondly, the use of radiation pressure from a low-NA beam in aqueous media is accompanied by absorption-mediated thermal effects, which can produce responses that may be an order of magnitude larger than the mechanical response induced by radiation-pressure force (Fig. 1a-d). Employing a differential scattering approach with 3-µm and 0.1-µm beads, our experiments in agarose hydrogels demonstrate that PF-OCE is able to isolate bead mechanical responses induced by modulated radiation pressure by compensating for the photothermal responses of the surrounding hydrogels. The measured total response from the 3-µm beads did not directly correspond to either the mechanical response or the photothermal response alone. These results confirm the importance of accounting for both scattering-mediated radiation pressure and absorption-mediated photothermal responses when using 976-nm photonic excitation from a low-NA beam in aqueous media.

One potential area of application of PF-OCE is the study of dynamic cell-ECM biophysical interactions in 3D environment. Particularly, the embedded scattering beads used in PF-OCE may also readily serve as the fiducial beads in TFM[46]. Recent innovations in TFM strive to extend traditional quasi-static and 2D dynamic studies to 3D volumetric measurements of dynamic cell-ECM interactions including collective (emergent) cellular behaviours[24,29,30,46]. An important capability that would enable such studies is the characterization of mechanical properties of hydrogel substrates at sub-cellular length scales over a millimetre-scale volume at timescales of minutes-to-hours. OT-AMR is currently still the leading technique for characterization of ECM mechanical properties during live cell imaging studies[35]. However, a practical challenge that persists in current OT-AMR studies is the precise alignment of the trapping beam to the centre of the marker beads to within 0.1 µm, which must be achieved on each individual bead being probed at a given time. This results in a net measurement time of ~8 s per bead achieved by recent state-of-the-art OT-AMR studies[35]. Furthermore, the detection of transverse bead displacements (typically on the order of $10$-$10^2$ nm) in OT-AMR is implemented via a trans-illumination geometry, which limits the sample thickness (and turbidity) that can be imaged. In this respect, PF-OCE has the benefit of epi-illumination, continuous-scanning acquisition that probes multiple beads at various depths in each B-scan, owing to the use of a low-NA beam to exert transversely-localized axial radiation-pressure force over an extended depth range.

Moving towards future implementation of PF-OCE in 3D live-cell imaging studies, the acquisition scheme presented here can be expanded to accommodate larger-scale volumetric acquisition over timescales of minutes-to-hours. For instance, consider a 3D multi-cellular imaging study over a volumetric FOV of 1 mm × 1 mm × 200 µm (depth range about the PF beam focal plane with viable PF-OCE mechanical response data) with statistical spatial sampling determined by a 15-µm average edge-to-edge bead separation (corresponding to ~48,500 beads within the volume). A feasible PF-OCE acquisition scheme may involve acquiring 2 volumes serially, each consisting of 3,000 BM-mode frames per slow-axis position (achieving ~140 pm sensitivity instead of the current 75 pm) and a frame rate of 200 Hz over a fast-axis scan range of 500 µm (the largest scan range achievable by our current galvanometer at this frame rate). At the transverse spatial sampling of 1 µm/pixel, the total PF-OCE acquisition time for the entire volume would be 8.33 hours, whereas and OT-AMR measurement (at 8 s per bead, assuming that volumetric measurements can be conducted up to 200 um depth) would take up to 4.5 days. If we were to use a resonant scanner[57] and acquire the same PF-OCE dataset at the frame rate of 2 kHz over a fast-axis scan range of 1 mm, the total acquisition time would reduce to only 25 minutes. In this case, PF-OCE offers over 2 orders of magnitude (>250 times) improvement in volumetric throughput over the current state-of-the-art OT-based microrheology techniques. Future studies will also address the quantitative reconstruction the local complex shear modulus from the measured PF-OCE bead mechanical responses by utilizing OCT-based depth-resolved measurement of radiation-pressure force[62]. Based on our differential scattering approach, there is also a future possibility of an all-endogenous PF-OCE that relies on native scattering of different constituents in the biological sample. With the potential to reconstruct microscale viscoelastic properties of hydrogels from variations in complex mechanical responses of embedded beads over millimetre-scale volumes, PF-OCE may unlock new research directions in cell mechanics and mechanobiology. A potential example of this is the study of how micro-mechanical properties and biophysical cell-extracellular matrix (ECM) interactions impact collective behaviour[29-31], such as the 'emergent' 3D migration patterns of invasive cancer cells.

## Methods

### 1. Sample preparation

First, a sample chamber shown in Fig. 2a was fabricated from microscope glass coverslips (Electron Microscopy Sciences, 22x22 mm, #0) for each agarose hydrogel sample. The coverslips were bonded together by an RTV silicone adhesive (Permatex, 80050) as shown in Fig. 2a. The chamber was left to cure for 24 hours before use.

The four agarose hydrogel samples were made by mixing solid agarose polymer (Fisher Scientific, BP1423) with room-temperature distilled water at the concentrations of 0.2%, 0.3%, 0.4% and 0.5% (w/w). The mixture was repeatedly heated in a microwave oven for 5 seconds then stirred for 10 seconds until all visible agarose solid had dissolved and the mixture became clear. Throughout the process, the total weight of the mixture was constantly checked for any loss of water to evaporation; distilled water was added accordingly. 3-µm polystyrene microsphere suspension (Sigma-Aldrich, LB30) was added to the dissolved mixture at the concentration of 6 µL/mL to achieve 15-µm mean edge-to-edge particle spacing. Then, 0.1-µm polystyrene microsphere suspension (Sigma-Aldrich, LB1) was added at a concentration of 0.12 µL/mL to achieve 2-µm mean edge-to-edge particle spacing. The mixture was stirred by hand to ensure all particles were dispersed before injecting into the pre-made glass chamber.

Finally, the glass chamber was sealed on both opening ends by a liquid sealing glue (Bob Smith Industries, Insta-Cure+). We found that it was crucial to carefully seal the open ends of the chamber to produce a confined container. When the open ends of the glass chamber were not sealed, the hydrogels underwent drastic structural and compositional change due to evaporation of water and the motion of the beads was apparent under the OCT system during data acquisition.

Separate hydrogel samples were made in aluminum trays (The Lab Depot, TLDD43-100) from the same agarose-distilled water mixture for rheometer testing. The samples for the rheometer testing were cut into a disk with diameter of 40 mm and thickness of 2 mm.

**2. Optical setup**

The optical setup (Fig. 1a) for measuring complex OPL oscillation consisted of an SD-OCT system with a broadband superluminescent diode (Thorlabs, LS2000B) with centre wavelength of 1,300 nm. The OCT beam focused with an NA of 0.14 and had transverse and axial resolutions of 4.5 μm and 3.7 μm in air, respectively (we note that, unlike confocal microscopy, OCT does not need high NA to achieve cellular resolution[63]). Combined in free-space with the sample arm of the OCT system is a laser diode (RPMC Lasers, R0976SB0500P) at the wavelength of 976 nm (in vacuum) acting as the PF forcing beam. A beam control module was used to optimize the PF forcing beam and ensure that it focused to the same position in 3D space as the OCT beam after going through the same OCT sample arm objective lenses (refer to Supplementary Method 3). The waist radius of the PF forcing beam was measured to be 3.19 μm at the focal plane, corresponding to an NA ~ 0.1 (Supplementary Figure 1).

**3. Data acquisition**

To maximize the force exerted by the PF forcing beam during acquisition, the alignment between the OCT beam and the PF forcing beam was checked at the beginning of each experiment (Supplementary Method). A 3D BM-mode acquisition scheme (Fig. 2b) was adopted. A 3D volume consisted of 10 BM-mode datasets, acquired at 10 slow-axis positions along the y-axis. Each BM-mode dataset consisted of 6,144 frames (200 frames/second) with 256 A-scans per frame. This acquisition scheme provides a transverse field-of-view of 200 μm × 10 μm at the spatial sampling density of 0.8 µm/pixel along the fast x-axis and 1 µm/pixel along the slow y-axis. Each spatial voxel contains 6,144 measurements of the sample response over time. During the acquisition of each BM-mode dataset, the PF forcing beam power was modulated by a function generator (Tektronix, AFG3051C), which sent a continuous 20-Hz sinusoidal modulation waveform to the laser diode controller. Since the modulation was provided externally by a function generator, asynchronous to the OCT acquisition control, we measured the function generator output at the $m$th A-scan in each frame to reconstruct the full PF drive waveform. The function generator voltage at the $m$th A-scan reflected the real part of the complex drive waveform, $\tilde{V}(t)$, at that A-scan. From this measurement, we calculated the phase of the drive waveform at the 1st acquired A-scan from $\varphi_{\text{drive}} = \varphi_{\text{m}} - \omega t_m$, where $\varphi_{\text{m}}$ and $t_m$ denote the phase and time at the $m$th A-scan, respectively. Then, we reconstructed the complex drive waveform as $\tilde{V}(t) = V_0 \exp(i(\omega t + \varphi_{\text{drive}}))$, where $V_0$ denotes the modulation amplitude. At the frame rate of 200 Hz, the OPL oscillation due to the 20-Hz modulation was sampled at 10 distinct phases per modulation cycle. Three such volumes were acquired at different regions in each sample. All synchronization and instrument controls were accomplished via a custom LabVIEW software.

**4. Data processing**

Data processing was implemented in MATLAB. The spatial-domain OCT image was reconstructed with standard procedures (background subtraction, spectrum resampling, dispersion correction, and inverse Fourier transformation). In order to efficiently process large 3D BM-mode datasets, only the depths containing the sample (601 pixels in depth out of 2048 acquired) was reconstructed; this was implemented with a high-speed SD-OCT processing method for depth-selective reconstruction[46,64]. The reconstructed spatial domain complex OCT image was first segmented into a photothermal data region (corresponding to the 0.1-µm beads and bottom glass surface of the sample chamber) and a total response data region (corresponding to the 3-µm beads) via thresholds based on magnitude of the reconstructed OCT image and OCT SNR (Fig. 3). The thresholds used to generate the results here are defined in Supplementary Table 2. Unless stated otherwise, all remaining processing steps outlined in this section were performed independently on each spatial pixel in the 3D OCT image that passed the thresholds, which encoded OPL oscillation resulting from 20-Hz modulation of the PF forcing beam power, acquired over 6,144 BM-mode frames.

The OPL oscillations due to the modulated PF forcing beam were estimated with a previously described method to reconstruct complex sample displacement in phase-sensitive OCE[65]. Briefly, the complex phase differences were calculated between every adjacent BM-mode frame at each spatial pixel. The complex phase differences, expressed as $e^{i\Delta\Phi(\mathbf{r},t)}$, were first registered to that of the top glass surface of the sample chamber to remove systematic noise and phase drifts across BM-mode frames, then filtered by a median filter (3×3 kernel, applied separately to the real and imaginary parts of $e^{i\Delta\Phi(\mathbf{r},t)}$). The real-valued phase differences, $\Delta\Phi(\mathbf{r},t)$, were obtained from the phase angle of $e^{i\Delta\Phi(\mathbf{r},t)}$, then, filtered by a Butterworth bandpass filter (±1 Hz pass-band centred around 20 Hz). The complex OPL oscillation at each spatial pixel was obtained after cumulative summation (integration in time) and Hilbert transformation of $\Delta\Phi(\mathbf{r},t)$ along the fast x-axis. This complex OPL oscillation corresponded to the raw photothermal response, $\Delta\text{OPL}_{\text{PT}}(x,y,z,t)$, and total response, $\Delta\text{OPL}_{\text{tot}}(x_\text{b},y_\text{b},z_\text{b},t)$, for spatial pixels in the photothermal response data region and total response data region, respectively. The vector $(x_\text{b},y_\text{b},z_\text{b})$ refers the set of pixel coordinates of each spatial pixel composing the 3-μm beads. Note that we omit the argument $\omega$ included in equations (4a-c) because $\omega = 2\pi(20\text{ Hz})$ is implied for all PF-OCE measurements.

Amplitude and phase-shift w.r.t. $\varphi_{\text{drive}}$ of the raw $\Delta\text{OPL}_{\text{PT}}(x,y,z,t)$ in each 3D dataset were curve-fit as a function of depth by the theoretical curve obtained from simulation and by a cubic polynomial function, respectively. A cubic polynomial function was used to fit the phase data because the theoretical simulation did not to reproduce the non-zero and depth-dependent phase delay observed experimentally (Supplementary Figure 4). The curve-fitting was done by minimizing the weighted sum-square error (SSE). To accommodate for low-SNR data with large phase noise, the SSE calculation was weighted by the OCT SNR in each spatial pixel such that measurements with higher SNR were weighted more heavily. The weights, $W$, were given by

$$W = \begin{cases} 0 & ; \quad \text{SNR} < 3 \\ \dfrac{\text{SNR} - 3}{4} & ; \quad 3 \leq \text{SNR} < 7 . \\ 1 & ; \quad \text{SNR} \geq 7 \end{cases} \qquad (5)$$

For each 3D dataset, the best-fit lines for the amplitude and phase data yielded $A_{\text{PT}}(z)$ and $\varphi_{\text{PT}}(z)$, respectively. Then, the depth-dependent complex photothermal response, $\Delta\text{OPL}_{\text{PT}}(z,t)$, was obtained from $\Delta\text{OPL}_{\text{PT}}(z,t) = A_{\text{PT}}(z) e^{i(\omega t + \varphi_{\text{drive}} + \varphi_{\text{PT}}(z))}$.

The complex mechanical response at each spatial pixel that made up the 3-µm beads was obtained from $\Delta\text{OPL}_{\text{mech}}(x_b, y_b, z_b, t) = \Delta\text{OPL}_{\text{tot}}(x_b, y_b, z_b, t) - \Delta\text{OPL}_{\text{PT}}(z_b, t)$. The isolated $\Delta\text{OPL}_{\text{mech}}(x_b, y_b, z_b, t)$ was further filtered, via multiplication in by a brick-wall filter (±0.2 Hz pass-band) in the frequency domain, before its amplitude $A_{\text{mech}}(x_b, y_b, z_b)$ and phase $\varphi_{\text{mech}}(x_b, y_b, z_b)$ were extracted. In order to obtain $A_{\text{mech}}$, $\varphi_{\text{mech}}$, $A_{\text{PT}}$, $\varphi_{\text{PT}}$, $A_{\text{tot}}$, and $\varphi_{\text{tot}}$ for each of the 3-µm beads, the spatial pixels that belonged to the same 3-µm beads were identified and grouped together by their pixel coordinates $(x_b, y_b, z_b)$. Then, the median values of $A_{\text{mech}}$, $\varphi_{\text{mech}}$, $A_{\text{PT}}$, $\varphi_{\text{PT}}$, $A_{\text{tot}}$, and $\varphi_{\text{tot}}$ were calculated for each group, corresponding to the responses of each 3-µm bead. These median values were used to generate the maps of $A_{\text{mech}}$ and $\varphi_{\text{mech}}$ (Fig. 4b, c).

## 5. Statistical analysis

All statistical analysis was implemented in MATLAB. Two statistical tests were performed. Firstly, a Wilcoxon-type non-parametric test for ordered groups, proposed by Cuzick[66], was implemented to test the null hypothesis that that there was no statistically significant trend across the four agarose concentrations (i.e., the responses from the four samples were not ordered) against the alternative hypothesis that there was a statistically significant trend. The normalized test statistics, $z_C$, and the associated 2-sided p-value, $p_C$, are reported. The data were ordered such that a $z_C > 0$ indicates an increasing trend while a $z_C < 0$ indicates a decreasing trend. We considered a trend to be statistically significant if $p_C < 0.05$. Secondly, a multiple comparison based on Wilcoxon-type non-parametric Kruskal-Wallis test of variance was implemented to determine if there were statistically significant differences between measurements from any two agarose concentrations. The reported p-values reflect the significance of chi-squared ($\chi^2$) statistics on the group-adjusted (Bonferroni correction for multiple comparisons among groups) 2-sided pairwise comparison between two agarose concentrations. In both tests, rank-based non-parametric methods were chosen to accommodate for deviation from a normal distribution (Anderson-Darling test for normality) and unequal variances (Barlett's test for equal variances) among measurements in different agarose concentrations.

Although the variations in the local mechanical responses within a sample can be attributed to the heterogeneity of agarose hydrogels (see Supplementary Discussion 2 for possible sources of variability within a sample), in order to compare these responses to the bulk mechanical properties measured via rheometry and to observe the overall trend of the mechanical response across different concentrations, the reconstructed data from Method 4 (i.e., $A_{\text{tot}}(x_b, y_b, z_b)$, $\varphi_{\text{tot}}(x_b, y_b, z_b)$, $A_{\text{mech}}(x_b, y_b, z_b)$, $\varphi_{\text{mech}}(x_b, y_b, z_b)$, $A_{\text{PT}}(z_b)$, and $\varphi_{\text{PT}}(z_b)$) were subjected to further thresholding to exclude outliers from the responses of each agarose concentration prior to performing the statistical tests. Any spatial pixels that contained $A_{\text{mech}}(x_b, y_b, z_b)$ values above the 85th percentile or below the 15th percentile of all $A_{\text{mech}}(x_b, y_b, z_b)$ values in each 3D dataset (i.e., the percentiles were calculated separately for each region in each sample), were excluded from the statistical tests. The acceptance or rejection of the null hypothesis by the Cuzick's test for trend was not affected by this exclusion (Supplementary Figure 9). The responses of each of the 3-μm beads were then obtained from the median values among all remaining spatial pixels in each 3D dataset that constitute each bead, as described in Method 4.

All statistical analysis, including all box plots (Fig. 5 and Supplementary Figures 6 & 7), were performed using the median values for each of the 3-μm beads after exclusion of outlier pixels. The total number of beads included in the statistical analysis was N = 54, 74, 58, and 52 beads for 0.2%, 0.3%, 0.4%, and 0.5% agarose hydrogel samples, respectively.

## Acknowledgement

The authors gratefully acknowledge useful discussion with Dr. Nozomi Nishimura regarding statistical analysis. This work was funded in part by National Institutes of Health (NIBIB-R21EB022927) and Cornell Discovery and Innovation Research Seed award to SGA. This work made use of the Cornell Center for Materials Research Shared Facilities which are supported through the NSF MRSEC program (DMR-1719875).

## Author contributions

NL and RRI set up and conducted PF-OCE experiments and theoretical simulations; NL conducted rheometry measurements; NL and RRI processed and analyzed all data; JAM developed the MATLAB code for depth-selective OCT reconstruction; GRU and JAM helped conceive theoretical simulations; GRU designed and implemented the LabVIEW acquisition software; SGA conceived the principle of PF-OCE and the experiments presented here; NL wrote this manuscript; SGA supervised the writing of this manuscript; All co-authors assisted in the revision of this manuscript.

## Competing financial interests

SGA, GRU and NL are listed as inventors on a patent related to the methods presented in this manuscript.

# Supplementary information for "Photonic force optical coherence elastography for three-dimensional mechanical microscopy"

Nichaluk Leartprapun[1], Rishyashring R. Iyer[1], Gavrielle R. Untracht[1,a], Jeffrey A. Mulligan[2], and Steven G. Adie[1,*]

[1] Meinig School of Biomedical Engineering, Cornell University, Ithaca, New York 14853

[2] School of Electrical and Computer Engineering, Cornell University, Ithaca, New York 14853

[a] Current affiliation: Optical and Biomedical Engineering Laboratory, School of Electrical, Electronic and Computer Engineering, The University of Western Australia, Perth, Western Australia 6009

* Corresponding author: sga42@cornell.edu

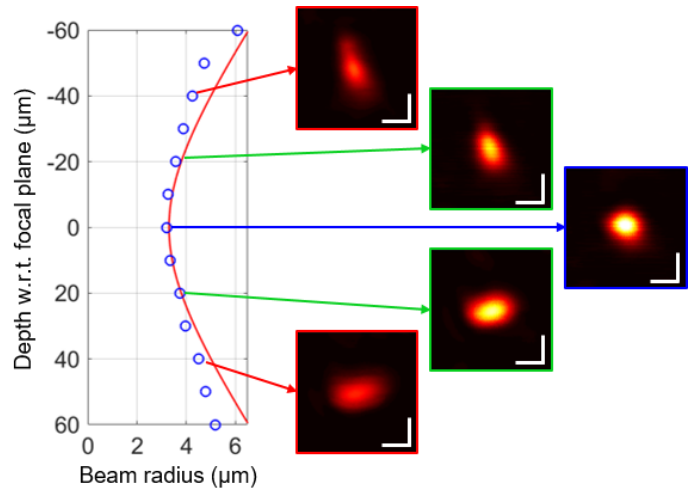

**Supplementary Figure 1. Point spread function of the PF forcing beam.** Plot shows comparison between the $1/e^2$ radii of the PSF obtained from the reflected confocal response of a single 0.5-μm polystyrene bead (**o**) and from the theoretical Gaussian beam profile with the same waist radius (—) as function of depth. Images show the *en face* views of the PSF measured at selected depths. Scale bar: 5 μm.

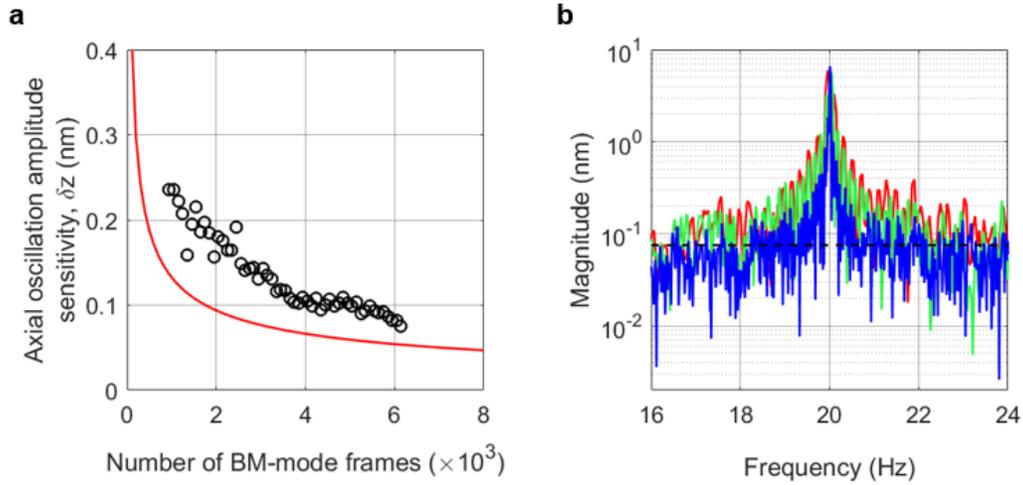

**Supplementary Figure 2. Dependence of OCT axial oscillation amplitude sensitivity on the number of BM-mode frames. a**, Observed (**o**) and shot-noise limited theoretical (—) axial oscillation amplitude sensitivity for an OCT signal with SNR of 25 dB as a function of the number of BM-mode frames. The observed oscillation amplitude noise floor with 6,144 BM-mode frames was 75 pm while the theoretical shot-noise limit was 54 pm. **b**, Power spectrum of measured $\Delta OPL_{tot}$, truncated to 6,144 (—), 4,044 (—) and 2,044 (—) frames. Dotted line indicates the 75-pm oscillation amplitude sensitivity. We note that, in practice, the smallest detectable oscillation amplitude is approximately 150 pm (3 dB above the noise floor).

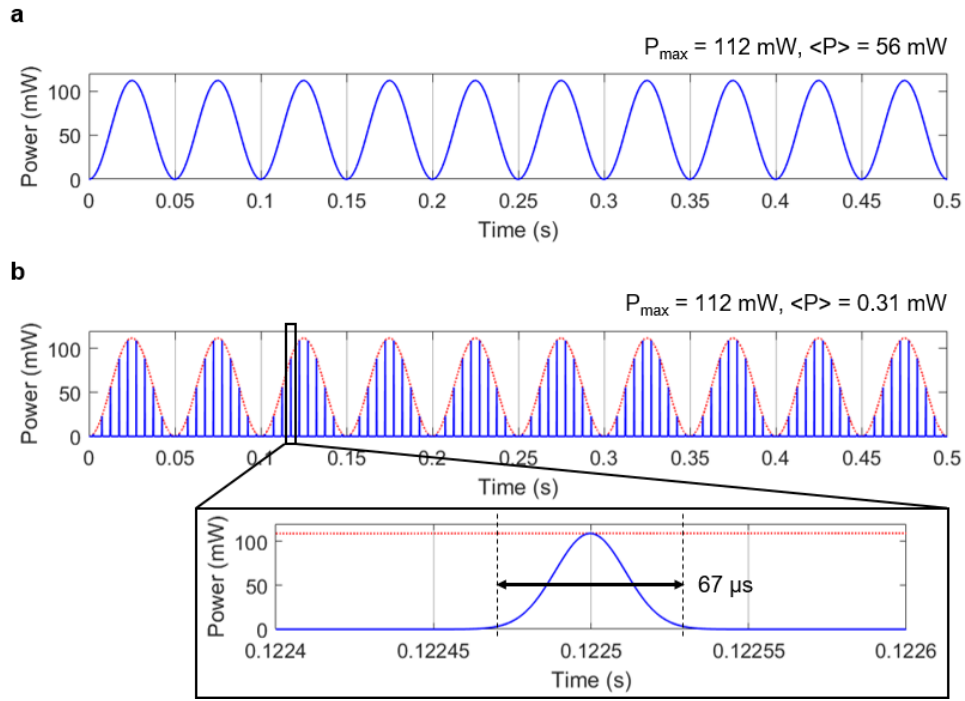

**Supplementary Figure 3. Comparison of continuous harmonic modulation versus pulse-train PF drive waveform. a**, Continuous 20-Hz sinusoidal drive waveform with peak power, $P_{max}$, of 112 mW generated by the function generator. **b**, Actual drive waveform felt by each of the 3-µm beads (—) due to beam-scanning along the fast axis in the BM-mode acquisition scheme resembled a pulse-train excitation with a 20-Hz sinusoidal envelope (····). This type of excitation resulted in 3 orders of magnitude lower time-average power, $\langle P \rangle$, compared to the continuous excitation case. The zoomed-up panel shows the pulse width of each pulse excitation based on the dwell time of the PF forcing beam on each 3-µm bead.

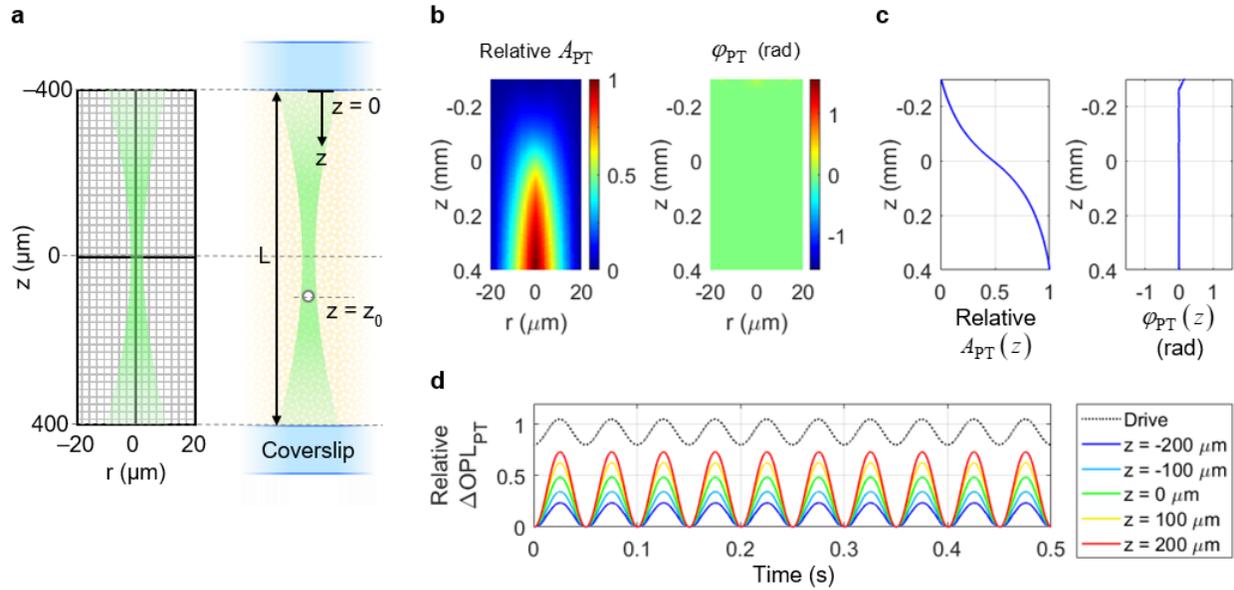

**Supplementary Figure 4. Illustrations of geometries used in theoretical simulation of photothermal response (Supplementary Method 1) and representative simulation output. a**, Space-domain in cylindrical coordinate, assuming symmetry about $r=0$ axis, for numerical integration of the heat transfer equation (Supplementary Equation 1). **b**, Geometry and dimensions used for cumulative $\Delta\mathrm{OPL}_{\mathrm{PT}}$ calculation by our modified version of Lapierre's model (Supplementary Equation 2). **c**, Maps of relative $A_{\mathrm{PT}}$ and $\varphi_{\mathrm{PT}}$ from the simulation using a beam with $\lambda = 976$ nm and $w_0 = 3.19$ μm showed spatially varying amplitude but uniformly zero phase delay. **d**, Depth-dependent curves taken at $r=0$ for relative $A_{\mathrm{PT}}(z)$ and $\varphi_{\mathrm{PT}}(z)$. **e**, Time profile of the real part of relative $\Delta\mathrm{OPL}_{\mathrm{PT}}$ taken from various depths at $r=0$. Response at all depths appeared to be in-phase with the drive waveform.

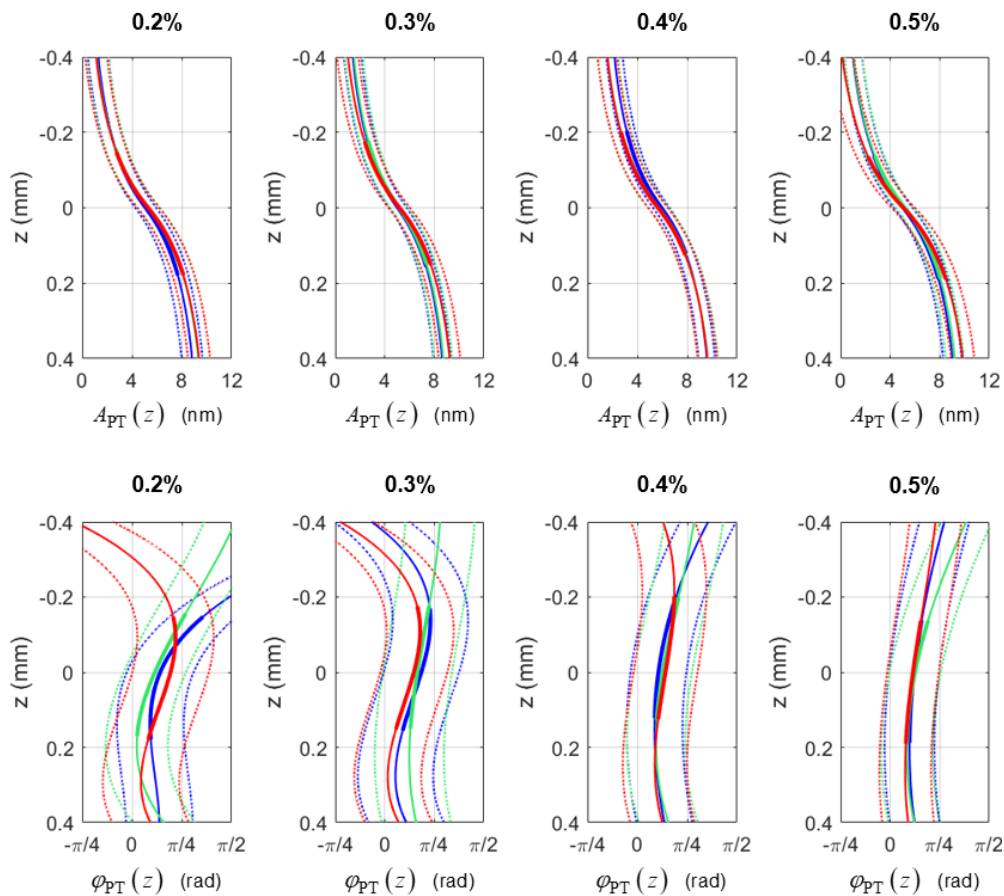

**Supplementary Figure 5. Experimental curve fits for the depth-dependent photothermal response.** Amplitude (top panels) and phase delay (bottom panels) of depth-dependent photothermal responses obtained from 3 regions (red, green, and blue lines) in each sample. Bolded lines indicate depth range around focal plane where PF data regions are located. —: best-fitted lines. ⋯: ±1 standard deviations. Although the amplitude curves were consistent across different sample regions and concentration, the phase delay curves were more variable with larger relative uncertainties. Particularly, the discrepancies across 3 regions within the same sample appeared to be more prominent at lower agarose concentrations. This could be a result of a larger degree of syneresis (dynamic fluid flow through the agarose polymer matrix, causing structural change due to gel swelling-deswelling over time) at lower agarose concentrations[1-3] as well as larger contributions of apparent photothermal response due to diffusive motions of the 0.1-μm beads in hydrogels with larger pores[4].

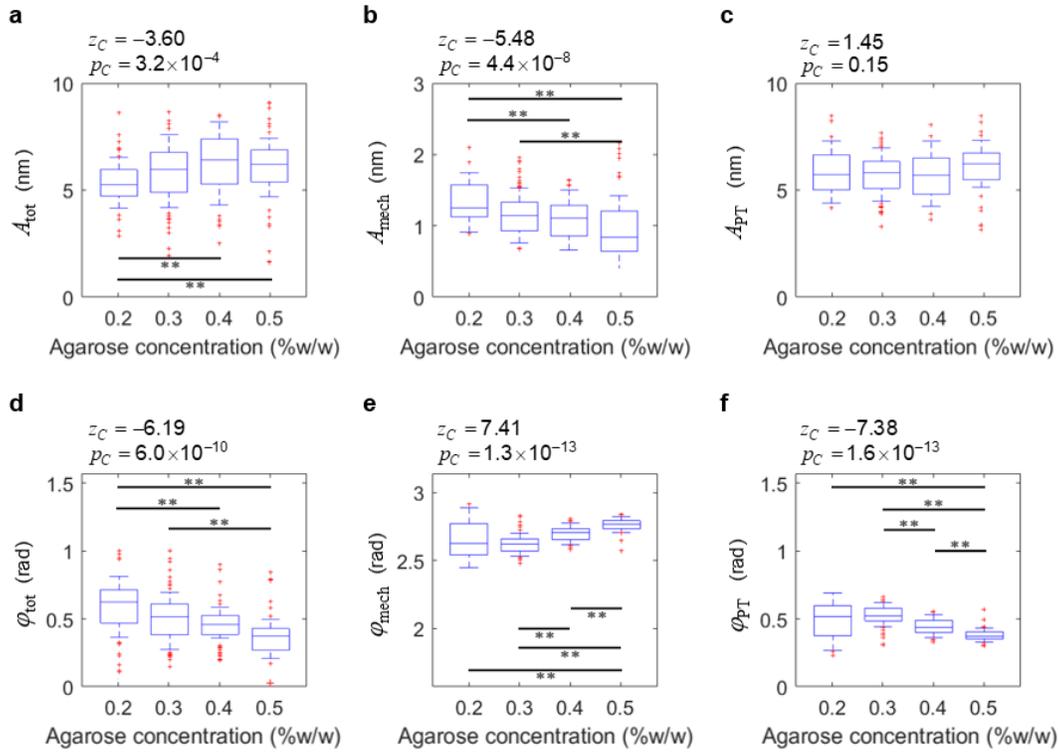

**Supplementary Figure 6. Box plots comparing complex total, mechanical and photothermal responses in agarose hydrogels.** Magnitude and phase of **a**, **d**, total response (omitted in Fig. 5), **b**, **e**, mechanical response, and **c**, **f**, photothermal response measured by PF-OCE. Horizontal lines within boxes indicate median values, boxes denote interquartile ranges. Error bars span one standard deviation; data outside of this range are shown in red markers. Black bar and asterisks indicate a statistically significant difference between 2 agarose concentrations per Kruskal-Wallis test at $p < 0.05$ (*) and $p < 0.005$ (**) confidence levels. $z_C$ and $p_C$, respectively, denote normalized test statistic and associated p-value for Cuzick's test for trend across the four agarose concentrations, ordered from 0.2% to 0.5% w/w; thus, $z_C > 0$ indicates an increasing trend while $z_C < 0$ indicates a decreasing trend. A trend was considered statistically significant if $p_C < 0.05$. Refer to Method 5 for details of the statistical analysis.

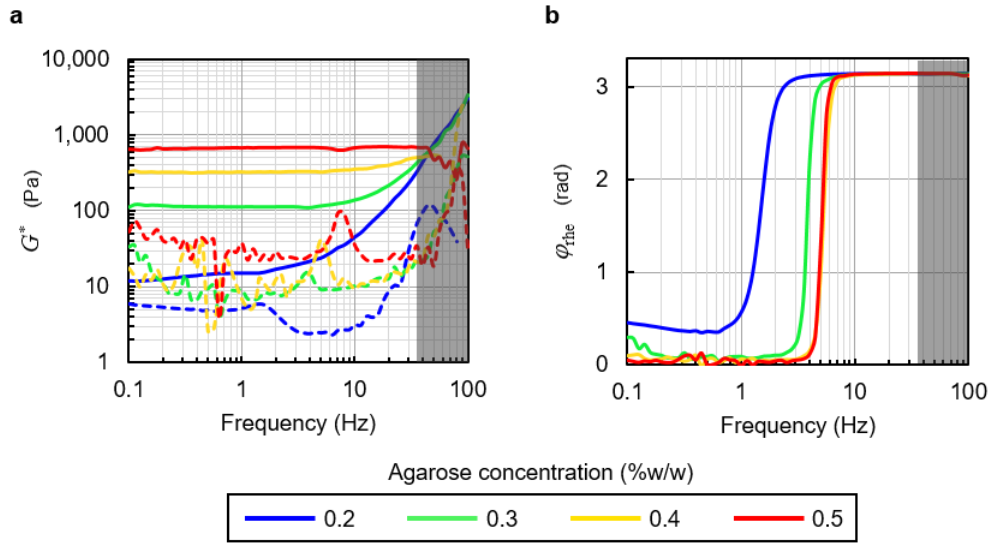

**Supplementary Figure 7. Complex shear modulus, $G^*$, and phase delay, $\varphi_{\text{rhe}}$, measured by oscillatory test on a parallel-plate shear rheometer. a**, Storage moduli (—) and loss moduli (---) as a function of oscillation frequency. **b**, Raw phase delays between displacement and applied torque. The frequencies at which the sharp rise in phase delays occur roughly correspond to the damped natural frequencies of the samples. Shaded regions correspond to the frequency range over which measurements were unreliable.

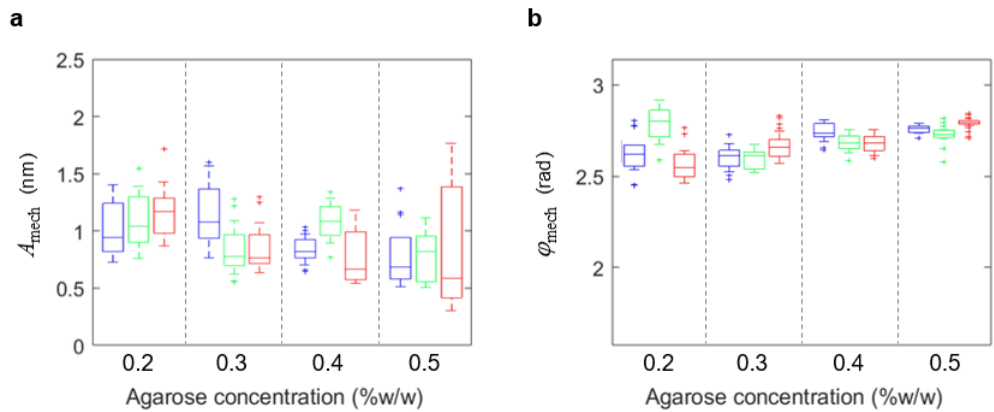

**Supplementary Figure 8. Comparison of bead mechanical responses from three regions in each hydrogel sample.** Box plots of **a**, amplitude and **b**, phase of bead mechanical responses measured from 3 regions (indicated by blue, green, and red colors) in each sample. Horizontal lines within boxes indicate median values, boxes denote interquartile ranges. Error bars span one standard deviation; data outside of this range are shown in red markers

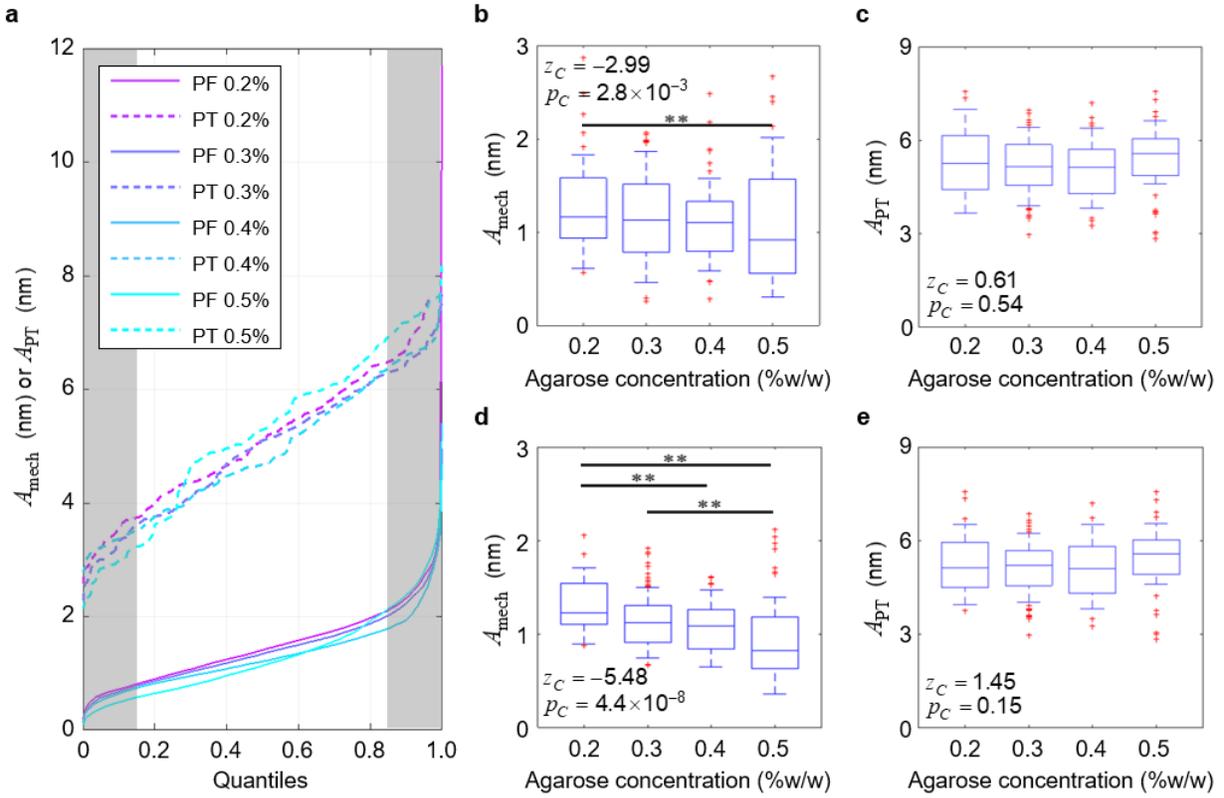

**Supplementary Figure 9. Exclusion of outliers for multiple comparisons by Kruskal-Wallis test among four agarose concentrations. a**, Quantiles of magnitude of mechanical and photothermal responses measured from each agarose concentration. The dramatic rise in $A_{\text{mech}}(x_b, y_b, z_b)$ around the lower and the higher quantiles (shaded regions) may reflect sources of variability described in Supplementary Discussion 2, and are considered as outliers for each concentration. However, from 0.15 to 0.85 quantiles, the $A_{\text{mech}}(x_b, y_b, z_b)$ curves maintain a monotonically increasing linear regime. In contrast, $A_{\text{PT}}(z_b)$ curves do not maintain any particular trend versus concentration throughout the entire distribution. The effect of excluding data in the shaded regions (outliers) is visualized by comparing magnitude of mechanical and photothermal responses **b**, **c**, before against **d**, **e**, after the exclusion of data. The general decreasing trend versus concentration is observed for $A_{\text{mech}}(x_b, y_b, z_b)$ in both cases, but the contrast between concentrations is more apparent after excluding the outliers from the analysis. No significant trend or difference between concentrations could be observed for $A_{\text{PT}}(z_b)$ in either case. In **b-e**, horizontal lines within boxes indicate median values, boxes denote interquartile ranges. Error bars span one standard deviation; data outside of this range are shown in red markers. Black bar and asterisks indicate a statistically significant difference between 2 agarose concentrations per Kruskal-Wallis test at $p < 0.05$ (*) and $p < 0.005$ (**) confidence levels. $z_C$

and $p_C$, respectively, denote normalized test statistic and associated p-value for Cuzick's test for trend across the four agarose concentrations, ordered from 0.2% to 0.5% w/w; thus, $z_C > 0$ indicates an increasing trend while $z_C < 0$ indicates a decreasing trend. A trend was considered statistically significant if $p_C < 0.05$. Refer to Method 5 for details of the statistical analysis.

**Supplementary Table 1.** Properties of agarose hydrogels.

| Physical properties | |
|---|---|
| Mass density, $\rho$ (kg m$^{-3}$)[5] ** | 1000 |
| Thermal conductivity, $k$ (W m$^{-1}$ K$^{-1}$)[6] * | 0.570 (0.2%) |
| | 0.554 (0.3%) |
| | 0.538 (0.4%) |
| | 0.523 (0.5%) |
| Specific heat capacity, $c_V$ (J kg$^{-1}$ K$^{-1}$)[5] ** | 4187 |

| Photothermal properties | |
|---|---|
| Absorption coefficient, $\alpha$ (m$^{-1}$)[7] ** | 50 (976 nm) |
| Thermo-optic coefficient, $dn/dT$ (K$^{-1}$)[8] ** | –9.17 × 10-5 |
| Thermal expansion coefficient, $\beta$ (m m$^{-1}$ K$^{-1}$)[9] * | 3.68 × 10-6 |

* Values were obtained from extrapolation of the reported results at higher agarose concentrations.

** Properties not found for agarose hydrogels, values taken for water instead.

**Supplementary Table 2.** Thresholds for OCT image segmentation.

| Notation | Description | Value used * |
|---|---|---|
| **Thresholds for 0.1-µm beads in photothermal data region** | | |
| $lb_{PT}$ | Lower bound of reconstructed OCT signal magnitude | 700 (a.u.) |
| $ub_{PT}$ | Upper bound of reconstructed OCT signal magnitude | 1400 (a.u.) |
| $lb_{PT,\,dB}$ | Lower bound of OCT SNR | 3 (dB) |
| $ub_{PT,\,dB}$ | Upper bound of OCT SNR | 12 (dB) |
| **Thresholds for 3-µm beads in total response data region** | | |
| $lb_{tot}$ | Lower bound of reconstructed OCT signal magnitude | 5,000 (a.u.) |
| $lb_{tot,\,dB}$ | Lower bound of OCT SNR | 25 (dB) |

* Values are specific to the results presented in this manuscript. Values may change according to different acquisition settings or experimental conditions.

**Supplementary Method 1: Theoretical simulation of photothermal response**

The parameters and the geometry used in the simulation can be found in Supplementary Table 1 and Supplementary Figure 4, respectively. First, the 3D heat transfer equation was numerically solved to obtain the change in temperature due to optical absorption by water molecules in the agarose hydrogels. We assumed cylindrical symmetry about the optical axis of the PF forcing beam and assumed zero heat flux normal to each of the four domain boundaries. We also assumed that the PF forcing beam was Gaussian, described by the waist radius obtained from the PSF measurement shown in Supplementary Figure 1. The differential equation is given by

$$\rho c_V \frac{\partial T}{\partial t} = k \left( \frac{1}{r} \frac{\partial}{\partial r} \left( r \frac{\partial T}{\partial r} \right) + \frac{\partial^2 T}{\partial z^2} \right) + P(r,z,t), \tag{1a}$$

and

$$P(r,z,t) = \frac{\alpha P_0}{\pi w^2(z)} e^{-\left( \frac{r^2}{w^2(z)} + \alpha z \right)} e^{i(\omega t + \varphi_{\text{drive}})}, \tag{1b}$$

where $r = [-20, 20]$ μm and $z = [-400, 400]$ μm denote the radial and axial coordinates, respectively, $T$ denotes temperature, $w$ denotes beam radius, $P_0$ denotes beam power and $\omega$ denotes the angular modulation frequency. As such, $P(r,z,t)$ denotes the absorbed optical power per unit volume. After solving for the temperature change, $\Delta T$, the OPL change due to photothermal response, $\Delta \text{OPL}_{\text{PT}}$, was calculated at each time step by

$$\Delta \text{OPL}_{\text{PT}} = \Delta \text{OPL}_{\text{above}} - \Delta \text{OPL}_{\text{below}} \tag{2a}$$

where

$$\Delta \text{OPL}_{\text{above}} = \int_0^{z_0} \left( n_0 + \frac{dn}{dT} \Delta T(r,z) \right) \left( 1 + \beta \Delta T(r,z) \right) dz - \int_0^{z_0} n_0 dz, \tag{2b}$$

and

$$\Delta \text{OPL}_{\text{below}} = \left[ \int_{z_0}^{L} (1 + \beta \Delta T(r,z)) dz - \int_{z_0}^{L} dz \right] \left( n_0 + \frac{dn}{dT} \Delta T(r, z_0) \right). \tag{2c}$$

The expressions $\Delta\text{OPL}_\text{above}$ and $\Delta\text{OPL}_\text{below}$ describe the OPL change measured on the bead due to the photothermal response of the portion of the medium located above the bead ($z \leq z_0$) and below the bead ($z_0 < z \leq L$), respectively. This geometry is illustrated in Supplementary Figure 4. We note that the model by Lapierre et al.[10] did not take into account $\Delta\text{OPL}_\text{below}$ and $\Delta\text{OPL}_\text{PT}$ was simply given by equation (2b). The measurements by Lapierre et al.[10] were acquired in a phantom with absorbing particles embedded in a less absorbing medium inside a large unconfined container, which allowed the medium surrounding the particles to expand freely due to the heating of the absorbing particles, and the OPL change depended only on the photothermal response of the medium above the bead. In other words, $\Delta\text{OPL}_\text{PT} = \Delta\text{OPL}_\text{above}$ when the thermal expansion of the medium located below the bead had negligible effect on the physical displacement of the bead relative to the top of the sample ($z = 0$). Here, we included the contribution of both the thermal expansion of the medium above the bead, which pushed the bead away from $z = 0$, in equation (2b) and the thermal expansion of the medium beneath the bead, which pushed the bead closer to $z = 0$, in equation (2c) to account for our thin confined sample configuration. In equation (2c), the terms inside the square brackets describe the physical displacement of the bead due to the thermal expansion of the medium below the bead. To obtain the OPL change, this physical displacement is multiplied by the refractive index of the medium where the bead is located, given by the remaining terms in equation (2c).

The initial refractive index of the medium, $n_0$, was assumed to be 1.4. We also used $P_0 = 1\,\text{W}$ in the simulation to generate large enough OPL change to ensure that the numerical solutions remained above the machine precision limit ($\sim 10^{-12}$). As a result, the magnitude of the simulated $\Delta\text{OPL}_\text{PT}$ could not be used for direct comparison against the experimentally measured $A_\text{PT}$. Alternatively, we normalized the magnitude of $\Delta\text{OPL}_\text{PT}$ by $|\Delta\text{OPL}_\text{PT}(r,z)|/|\Delta\text{OPL}_\text{PT}(0,L)|$ and obtained relative $A_\text{PT}$. See Supplementary Figure 4 for example of simulation output.

**Supplementary Method 2: Order-of-magnitude estimation of relative force and displacement of 3-µm and 0.1-µm beads**

In order to measure both the total response, $\Delta \text{OPL}_{\text{tot}}$, and the photothermal response, $\Delta \text{OPL}_{\text{PT}}$, we used two bead sizes: higher scattering 3-µm beads provided access to $\Delta \text{OPL}_{\text{tot}}$ and low-scattering 0.1-µm beads allowed us to isolate $\Delta \text{OPL}_{\text{tot}}$. The 3-µm beads were added to achieve a mean particle separation of 15 µm in 3D. The 0.1-µm beads were added at a much higher density, achieving a mean particle separation of only 2 µm. In order to justify our assumption that the OPL oscillation measured on the 0.1-µm beads reflected the photothermal response alone, with negligible contribution of the mechanical response, we considered the theoretical prediction of two extreme cases.

Theoretical prediction (weak mechanical coupling limit). We considered two single-bead scenarios (either 0.1-µm or 3-µm beads), neglecting the presence of multiple beads within the excitation volume. Essentially this assumes that the displacements of a bead are completely independent to those of nearby beads. In this case, the relative magnitude of radiation-pressure force between the 3-µm and the 0.1-µm beads can directly be obtained from separate GLMT simulations for each bead size. We obtained the ratio of radiation-pressure force magnitude per unit power for the two bead sizes of $\bar{F}_{\text{rad},3\mu m}/\bar{F}_{\text{rad},0.1\mu m} \sim 10^5$. Likewise, the resulting bead oscillation amplitude could be obtained from Oestreicher's model[11], independently for each bead size. We obtained the resulting ratio of bead oscillation amplitude per unit power of $\bar{u}_{0,3\mu m}/\bar{u}_{0,0.1\mu m} \sim 10^4$.

Theoretical predication (strong mechanical coupling limit). We considered an equivalent rigid body scenario and accounted for the presence of multiple 0.1-µm beads (or a separate scenario with a single 3-µm bead) within the excitation volume. We approximated the PF excitation volume as a rigid sphere with radius of 3.19 µm, the $1/e^2$ radius of our PF forcing beam. In the strong mechanical coupling limit (all beads within the excitation volume are connected by rigid rods), the net external force acting on the rigid PF excitation volume was the sum of all forces on all beads inside the excitation volume. Given our mean particle separation, the ratio of the number of beads inside the excitation volume was $N_{3\mu m}/N_{0.1\mu m} \sim 10^{-2}$. In this case, we obtained the net radiation-pressure force ratio on the PF excitation volume for the two scenarios of $\bar{F}_{\mathrm{rad},3\mu m}/\bar{F}_{\mathrm{rad},0.1\mu m} \sim 10^5 \cdot 10^{-2} = 10^3$. Since the excitation volume was a rigid sphere of the same size for both cases, we also obtained the resulting bead oscillation amplitude ratio of $\bar{u}_{0,3\mu m}/\bar{u}_{0,0.1\mu m} \sim 10^3$.

Experimental prediction based on OCT scattering. We expected that the actual relative difference in the oscillation amplitude of the 3-µm and 0.1-µm beads in our experiments would lie somewhere between these two extreme theoretical cases for weak or strong mechanical coupling. As an experimental comparison to our theoretical estimation, we used the observed OCT scattering intensity to infer the relative magnitude of radiation pressure on the two bead sizes. We note that this assumes that the relative backscattering from the 0.1-µm versus 3-µm beads at 976 nm is comparable to the relative backscattering collected by the OCT system from the 0.1-µm versus 3-µm beads at 1,300 nm. The ratio of observed maximum OCT scattering intensity for the two bead sizes was $\left|\tilde{S}_{3\mu m}\right|^2 / \left|\tilde{S}_{0.1\mu m}\right|^2 \sim 10^3$, where $\tilde{S}$ denotes the complex OCT signal. The OCT scattering intensity in each pixel of the OCT image included the contribution from all beads inside the point spread function (PSF) of the OCT beam. At the focal plane, the OCT beam had comparable waist radius to the PF forcing beam. Given our mean particle separation, the ratio of the number of beads inside the PSF of the OCT beam was $N_{3\mu m}/N_{0.1\mu m} \sim 10^{-2}$. Assuming all beads inside the PSF contributed equally to the observed OCT scattering intensity, we arrived at the radiation-pressure force per one 3-µm bead to radiation-pressure force per one 0.1-µm bead ratio of $\bar{F}_{\mathrm{rad},3\mu m}/\bar{F}_{\mathrm{rad},0.1\mu m} \sim 10^3/10^{-2} = 10^5$, which agrees with our theoretical prediction in the weak mechanical coupling limit.

**Supplementary Method 3: Characterization of the PF forcing beam**

The point spread function (PSF) of the PF forcing beam was characterized on a single 0.5-μm polystyrene bead on a monolayer phantom in air. A photoreceiver (Newport, 2051-FS) was used to collect the reflected confocal response from a bead as the monolayer phantom was translated to various depths. The monolayer phantom was made by first, diluting a stock solution of 0.5-μm polystyrene bead suspension (Sigma-Aldrich, LB5, 10% solids) in ethanol (VWR, Ethanol, Pure) to a 1:$10^9$ volume ratio. A 1-μL drop of the diluted microsphere solution was spread on top of an anti-reflection-coated plano-convex lens (Thorlabs, LA1213-B), then left to sit until all solvent had evaporated. The AR-coated plano-convex lens was necessary to sufficiently reduce background reflection detected by the photoreceiver. The PSF of the PF forcing beam used in the experiments in this paper is shown in Supplementary Figure 1.

**Supplementary Method 4: Beam alignment procedure**

To maximize the force exerted by the PF forcing beam during the acquisition, the PF forcing beam was aligned to the OCT imaging beam such that the two beams focused to the same position in 3D space. This alignment was checked before every experiment. The position of the PF forcing beam was adjusted by the beam control module (BCM) consisting of six components: two collimating lenses, two spherical-aberration-compensation lenses and two right-angle mirrors. A photoreceiver (Newport, 2051-FS) was used to detect the reflected confocal response of the PF forcing beam. First, the alignment in the axial direction (along the optical axis of the two beams) was done by imaging a flat glass slide. The OCT focal plane was located by translating the glass slide in the axial direction and identifying the depth at which the detected OCT signal from the glass surface was maximized. The PF forcing beam focal plane was similarly located by identifying the depth at which the intensity detected by the photoreceiver was maximized. The position of the collimating lenses in the BCM were adjusted until the focal planes of the two beams were coplanar. Next, the alignment in the transverse plane was determined by imaging a USAF target. The confocal image from the PF forcing beam was compared to the *en face* view of the 3D OCT image. The right-angle mirrors in the BCM were tipped and tilted to steer the PF forcing beam in the transverse plane until both images of the USAF target were aligned.

**Supplementary Method 5: Calculation of OCT SNR and shot noise-limited oscillation amplitude sensitivity**

The oscillation amplitude sensitivity, fundamentally limited by the OCT phase noise floor, specifies the best achievable precision of the OPL oscillation measurements. In the shot noise limit, the smallest detectable phase difference between two adjacent BM-mode frames (i.e. phase difference at a particular spatial pixel at two time points), $\delta\Delta\phi$, depends on the SNR of the OCT signal and is given by Park et al.[12]

$$\delta\Delta\phi = \frac{1}{\sqrt{\mathrm{SNR}}}. \tag{3a}$$

Then, the physical (not OPL) oscillation amplitude sensitivity, $\delta z$, obtained from the reconstructed OPL oscillation (a series of phase difference measurements) is given, based on Chang et al.[13], by

$$\delta z = \frac{1}{\sqrt{\mathrm{SNR} \cdot N_A \cdot N_B}} \frac{\lambda}{4\pi n_{\mathrm{med}}}, \tag{3b}$$

where $n_{\mathrm{med}}$ denotes refractive index of the medium while $N_A$ and $N_B$ denote the number of samples per modulation cycle and the total number of modulation cycles measured, respectively. We note that the expression from Chang et al.[13] differs from Supplementary Equation (3b) by a factor of $1/\sqrt{2}$ because the authors considered the smallest detectable phase, $\delta\phi = 1/\sqrt{2 \cdot \mathrm{SNR}}$, whereas we considered the smallest detectable phase difference between two adjacent BM-mode frames, $\delta\Delta\phi$ [12], since this was what we used in our calculation of ΔOPL. For the 6,144-frame BM-mode acquisition scheme implemented in our experiments, $N_A$ was kept at 10 frames per modulation cycle (200-Hz frame rate at 20-Hz modulation frequency) and $N_B$ was 614 full modulation cycles per BM-mode dataset. The SNR was experimentally approximated by

$$\mathrm{SNR}(x, y, z_i) = \left( \frac{\sum_{i-2}^{i+2} \left| \tilde{S}_{\mathrm{signal}}(x, y, z_i) \right|}{\sum_{i-2}^{i+2} \left| \tilde{S}_{\mathrm{noise}}(z_i) \right|} \right)^2, \tag{4}$$

where $z_i$ denotes the pixel depth corresponding to the maximum OCT intensity on a particle, $\tilde{S}_{\text{signal}}(x, y, z)$ denotes the complex OCT signal after reconstruction, and $\tilde{S}_{\text{noise}}(z)$ denotes the complex depth-dependent noise floor of the OCT signal. $\tilde{S}_{\text{noise}}(z)$ was obtained from the difference of two reconstructed space-domain OCT background images, averaged across the transverse plane to obtain a depth-dependent complex OCT noise. The theoretical and observed oscillation amplitude sensitivity as a function of the number of BM-mode frames are shown in Supplementary Figure 2. The observed displacement sensitivity as a function of the number of BM-mode frames was obtained from the Fourier transform of the time-domain $\Delta \text{OPL}_{\text{tot}}$, truncated to a decreasing number of frames from the original 6,144-frame data. The oscillation amplitude noise floor was calculated from the root-mean-square of the noise spectrum adjacent to the 20-Hz response peak.

**Supplementary Method 6: Parallel-plate shear rheometer measurements**

A parallel-plate shear rheometer (TA Instruments DHR-3) with 40 mm diameter Peltier plate was used to measure the bulk complex shear modulus and phase delay of the agarose hydrogels. 3 samples were tested for each agarose concentration; each sample was tested 3 times consecutively. With these measurement conditions, we note that the rheometer testing does not replicate the boundary conditions presented in the PF-OCE measurements. Each test consisted of an oscillatory sweep from 0.1-100 Hz with torque amplitude of 10 µN·m. The complex shear modulus and phase delay from a representative test for each agarose concentration can be found in Supplementary Figure 5. Although the shear rheometer could operate up to 100-Hz oscillation frequency, results at higher frequencies (typically >20 Hz) are more prone to errors due to sample slippage and inaccurate phase angle measurements above the damped natural frequency (indicated by the sharp rise in phase delay). Any data points that reported negative values for either the real or the imaginary part of the complex shear modulus were excluded because they indicated inaccurate measurements.

**Supplementary Discussion 1: Reliability and accuracy of the theoretical simulation of the photothermal response.**

In principle, the contributions of both amplitude and phase of $\Delta \text{OPL}_{\text{PT}}$ in a homogeneous sample may be obtained from the theoretical simulation based on the model of photothermal phenomena in PT-OCT. However, we were not able to ascertain the accuracy of our theoretical photothermal response simulation under our experimental conditions due to several reasons. Firstly, many of the material properties required for the simulation were not available in the literature for our specific agarose hydrogels. We made approximations by taking properties of water or extrapolated from the available results in the literature to complete our simulation (Supplementary Table 1). Secondly, our current numerical simulation in MATLAB was limited by machine precision, in addition to the fact that we might not have accurately modelled the boundary conditions presented during the experiments. For instance, we were not able to account for the effects of the glass chamber, which could act as a heat sink that may affect the heat transfer process. Thirdly, our simulation might not have accurately accounted for all photothermal phenomena presented in the experiments. Given the similarities between our experimental conditions and those in existing PT-OCT studies, we adapted our model of OPL change due to absorption from PT-OCT literature. We accounted for two photothermal phenomena: the thermo-optic effect and thermal expansion. However, other photothermal phenomena such as photo-elastic effect[14], acousto-optic effects[15] and thermal forces[16,17] could also be present; we did not consider them to be significant photothermal contributions due to the absence of their discussion in PT-OCT literature. Further investigation into these photothermal phenomena will be needed to incorporate them into our simulation. Finally, our current simulation did not reproduce the phase delay we experimentally observed in the measured $\Delta \text{OPL}_{\text{PT}}$ from the 0.1-μm beads (Supplementary Figure 4). As a result, we could not use the theoretical simulation of $\Delta \text{OPL}_{\text{PT}}$ by itself to isolate the mechanical response. However, since the simulation produced the depth-dependent photothermal response amplitude that agreed with the general trends with observed on the 0.1-μm beads, we were able to use the functional form of the simulated $A_{\text{PT}}(z)$ curve to fit the experimental data.

**Supplementary Discussion 2: Sources of variability in PF-OCE measurements.**

Factors that may have contributed to the variability in the mechanical responses of the 3-µm beads in each agarose hydrogel sample include noise in OPL oscillation measurements, possible errors in compensating for the photothermal responses (which we have discussed in the main article), the BM-mode beam-scanning acquisition scheme, and microscale mechanical heterogeneity of the agarose hydrogels.

The accuracy of OPL oscillation measurements is dependent on the SNR of the complex OCT signal. This source of error is most prominent in the photothermal response measured from the 0.1-µm beads, which have low OCT SNR (0-12 dB) due to their weak scattering. However, under the assumption of transversely uniform photothermal response, the precision of measuring variations in bead mechanical responses within each sample is governed by the sensitivity of the OPL oscillation measurements on the 3-µm beads, which have higher SNR (>25 dB). Our observed oscillation amplitude sensitivity was approximately 75 pm (Supplementary Figure 2). Under the assumption of transversely uniform photothermal response (i.e., the error from the photothermal curve fits imposes a systematic error on the isolated mechanical response), this corresponds to the smallest practically detectable differences in $A_{\text{mech}}$ of various beads within a sample of 150 pm (3 dB above the noise floor) for any $A_{\text{mech}}$ values larger than 0.7 nm (the uncertainty of $A_{\text{PT}}(z)$ curves). In addition to OPL measurement errors and data-acquisition-related factors, Brownian motion of the beads inside porous structures of the agarose hydrogels may also confound the measurements of OPL oscillation induced by the PF forcing beam. It has been shown that Brownian motion of scattering particles resulted in a complex OCT signal with a Lorentzian distribution in the temporal frequency domain, which could also contribute to our observed displacement noise floor being above the shot noise limit.

As a result of the 3D BM-mode acquisition scheme the PF forcing beam was transversely aligned to the centre of each 3-µm bead for only a fraction of the measurements (approximately 10 out of 60 spatial pixels that constituted each bead). In other words, the magnitude of the radiation-pressure force exerted on a given 3-µm bead was varied as the PF forcing beam was scanned over different parts of each bead. Thus, we also expect that the resulting oscillations of the 3-µm beads could also be affected by this variation in the magnitude of force due to the scanning of the PF forcing beam.

Part of the variability observed in the maps of amplitude and phase of the 3-µm bead mechanical responses (Fig. 4b, c in the main manuscript) could also reflect the actual microscale heterogeneity that was present in the agarose hydrogels. Existing evidence of structural and compositional variability in agarose hydrogels support the notion that agarose hydrogels are mechanically heterogeneous at the microscale. Agarose hydrogel is composed of aggregated agar double-helix polymer matrix[18,19], creating a porous structure that holds water within its pores (diameter on the order of 1-6 µm for 0.2-0.5% agarose hydrogels)[4,20]. Additionally, the distribution of pore sizes of one concentration overlaps with that of others based on AFM measurements[20]. The porous structure could result in spatially heterogeneous mechanical responses measured by PF-OCE on each of the 3-µm beads. For instance, beads with larger $A_{\text{mech}}$ in each sample may be located inside larger pores, diffusing in the fluid phase of the biphasic hydrogels, whereas those with lower $A_{\text{mech}}$ may be trapped in the solid phase made up of agarose polymer matrix. Furthermore, the microstructure of agarose hydrogels is known to change over time due to multiple naturally occurring dynamic processes, including agarose fibre aggregation and local water expulsion from pores, collectively called 'syneresis'[1-3]. This dynamic change may have caused the variability between measurements from the different regions in each sample (Supplementary Figure 8), which were acquired at different times (separated by >1 hour) after the sample was first made.